\documentclass[useAMS,usenatbib]{mn2e}

\usepackage{epsfig}
\title[High resolution spectroscopy of IRAS~22023+5249]{High resolution spectroscopy of the high velocity hot post$-$AGB star LS~III~+52$^{\circ}$24 (IRAS~22023+5249)}
\author[G. Sarkar et al.]{G.~Sarkar$^{1}$\thanks{E-mail: gsarkar@iitk.ac.in}, D. A.~Garc\'{\i}a-Hern\'andez$^{2,3}$, M.~Parthasarathy$^{4,5}$, A.~Manchado$^{2,3,6}$\newauthor P.~Garc\'{\i}a-Lario$^{7}$ and Y.~Takeda$^{4}$\\
$^{1}$Department of Physics, Indian Institute of Technology, Kanpur$-$208016, U.P., India\\
$^{2}$Instituto de Astrof\'{\i}sica de Canarias, V\'{\i}a L\'actea s/n, E-38200 La Laguna, Spain\\
$^{3}$Departamento de Astrof\'{\i}sica, Universidad de La Laguna (ULL), E-38205 La Laguna, Spain\\
$^{4}$National Astronomical Observatory of Japan, 2-21-1 Osawa, Mitaka, Tokyo 181-8588, Japan\\
$^{5}$Aryabhatta Research Institute of Observational Sciences, Nainital, 263129,India\\
$^{6}$Consejo Superior de Investigaciones Cient\'{\i}ficas, Spain\\
$^{7}$Herschel Science Centre. European Space Astronomy Centre. Villafranca del Castillo, P.O. Box 50727. E-28080 Madrid. Spain}
\begin{document}

\date{Accepted xx xx xx. Received 2011 xx xx; in original form 2011 xx xx}

\pagerange{\pageref{page}--\pageref{lastpage}} \pubyear{2011}

\maketitle

\label{firstpage}

\begin{abstract} 
The first high-resolution (R$\sim$50,000) optical spectrum of the B$-$type
star,  LS~III~+52$^{\circ}$24, identified as the optical counterpart of the hot
post$-$AGB candidate IRAS~22023+5249 (I22023) is presented. We report the
detailed identifications of the observed absorption and emission features in the
full wavelength range (4290$-$9015 \AA) as well as the atmospheric parameters
and photospheric abundances (under the Local Thermodinamic Equilibrium
approximation) for the first time. The nebular parameters (T$_{\rm e}$, N$_{\rm
e}$) are also derived. We estimate T$_{\rm eff}$=24,000~K, log~g=3.0, $\xi_{\rm
t}$=7 kms$^{-1}$ and the derived  abundances indicate a slightly metal-deficient
evolved star with C/O$<$1. The observed P$-$Cygni profiles of hydrogen and
helium clearly indicate on-going post$-$AGB mass loss. The presence of [N~II]
and [S~II] lines and the non$-$detection of [O~III] indicate that
photoionisation has just started. The observed spectral features, large
heliocentric radial velocity, atmospheric parameters, and chemical composition
indicate that I22023 is an evolved post$-$AGB star belonging to the old disk
population. The derived nebular parameters (T$_{\rm e}$=7000~K, N$_{\rm
e}$=1.2$\times$10$^{4}$~cm$^{-3}$) also suggest that I22023 may be evolving into
a compact, young low-excitation Planetary Nebula. Our optical spectroscopic
analysis together with the recent Spitzer detection of double-dust chemistry
(the simultaneous presence of carbonaceous molecules and amorphous silicates) in
I22023 and other B-type post-AGB candidates may point to a binary system with a
dusty disk as the stellar origin common to the hot post-AGB stars with O-rich
central stars. 
\end{abstract}


\section{Introduction}

The discovery of high Galactic latitude cool (F, G, K) and hot (O,B) post
Asymptotic Giant Branch (post-AGB) supergiants - e.g., HD 161796 (Parthasarathy
\& Pottasch 1986) and LSII +34$^{\circ}$26 (Parthasarathy 1993) - indicated that
K, G, F, A, O, B post$-$AGB supergiants form an evolutionary sequence in the 
transition region from the tip of the AGB into the early stages of Planetary
Nebulae (PNe). Since then, several cool and hot post$-$AGB candidates have been 
identified (Pottasch \& Parthasarathy 1988; Parthasarathy \& Pottasch 1989; 
Garc\'ia$-$Lario et al. 1997a; Parthasarathy et al. 2000a). Gauba \&
Parthasarathy (2003, 2004) analysed the UV spectra and circumstellar dust
envelopes of several hot post$-$AGB stars from the above lists, including
LS~III~+52$^{\circ}$24. Yet to date, the high-resolution (R$\geq$30,000) optical
spectra of only a few hot post$-$AGB stars have been studied (Sarkar et al. 2005
and references therein). In order to unveil the evolutionary origins,
atmospheric parameters, and chemical compositions of hot post-AGB stars, there
is a clear need to carry out detailed spectroscopic studies of more examples of
this exotic class of post-AGB stars.

The optically bright B$-$type star, LS~III~+52$^{\circ}$24 identified with the
IR source IRAS~22023+5249 (hereafter, I22023) has far$-$IR colors similar to PNe
(see Table 1). It is listed in Wackerling's (1970) Catalog of early$-$type
emission$-$line stars. Recent ground$-$based high spatial resolution images in
the near-IR have shown H$_{2}$ emission arising close to the central star,
possibly in an incipient bipolar morphology (Volk et al. 2004). Indeed, Kelly \&
Hrivnak (2005) show that the excitation mechanism of the H$_{2}$ emission in
I22023 is a combination of radiative (fluorescence) and thermal (shock)
excitation. Gauba \& Parthasarathy (2004) reported the presence of weak
amorphous (10.8$\mu$m) and  crystalline (33.6$\mu$m) silicate features in the
I22023's Infrared Space Observatory (ISO) spectrum and classified the object as
O-rich. However, more recent and higher sensitiviy Spitzer/IRS spectra show that
I22023 display a mixed-chemistry (both C-rich and O-rich dust features) with the
presence of the classical aromatic infrared bands (AIBs; e.g., those at 6.2,
7.7, 8.6, and 11.3 $\mu$m) together with broad 10 $\mu$m amorphous silicate
emission and a strong IR excess (Cerrigone et al. 2009).

In this paper, we explore the first high-resolution (R$\sim$50,000) optical
spectrum of I22023 in order to unveil its evolutionary status and chemical
composition and to learn about the stellar origins of this peculiar type of hot
post-AGB stars. In Sect. 2 we briefly describe the optical observations of
I22023 and the data reduction process. A detailed analysis of the optical
spectrum is presented in Sect. 3 while the photospheric and nebular analysis
performed are shown in Sect. 4. We finish with a discussion and conclusions in
Sect. 5.

\section{Observations and data reduction}

I22023 was observed on 14 July 2001 using the Utrecht Echelle Spectrograph (UES)
on the 4.2m William Herschel Telescope (WHT) at  the Roque de los Muchachos
Observatory in La Palma (Spain). The observations were made with the 31.6
lines/mm echelle grating (E31), SITe1 CCD (2048 $\times$ 2048 pixels of 24
$\mu$m), a slit width of  1\arcsec on the sky and a central wavelength of
5500 \AA, resulting in a resolving power of R$\sim$50,000. The wavelength
coverage was 4290$-$4735 \AA~, 4760$-$5553 \AA~ and 5607$-$9015 \AA. A Th$-$Ar
comparison lamp was used for wavelength calibration. 

The one-dimensional spectrum was extracted using standard reduction procedures 
for echelle spectroscopy in the IRAF package. The data reduction steps included
bias and scattered light subtraction, flat-field correction, order extraction,
and wavelength calibration. The reduced spectrum was continuum$-$normalised. The
final signal-to-noise (S/N) ratio varied from 30 in the blue to more than 60
towards the red end of the spectrum. 

\begin{table*}
\centering
 \begin{minipage}{180mm}
  \caption{Details of the star}
\begin{tabular}{|c|c|c|c|c|c|c|c|c|c|c|c|c|}
\hline
IRAS & Name & RA & DEC & l & b & Sp. Type & V & B$-$V & \multicolumn{4}{c|}{IRAS Fluxes (Jy)} \\
     &      & 2000 & 2000 & &  &  Optical &   &       & 12 $\mu$m & 25 $\mu$m & 60 $\mu$m & 100 $\mu$m \\ 
\hline
\hline
22023+5249 & LSIII +52 24 & 22:04:12.30 & +53:04:01.4 & 99.30 & $-$1.96 & B$^{\rm a}$ & 12.52$^{\rm b}$ & 0.69$^{\rm b}$ & 1.02 & 24.69 & 14.52 & 3.93L \\
\hline
\end{tabular}
\indent \parbox{16cm}{$^{a}$Spectral type is from the SIMBAD database. $^{b}$Photometry is from Hog et al. (2000). 
L flag indicates that the quoted IRAS flux density is an upper limit.}
\end{minipage}
\end{table*}

\section{Analysis of the optical spectrum} 
Equivalent widths (W$_{\rm \lambda}$) of the absorption and emission lines were
measured. Deblending was done whenever required to obtain Gaussian fits to the
blended line profiles.  The complete continuum$-$normalised spectrum of  I22023
is presented in the appendix (Figure 4). This spectrum would be useful for
future observers since post$-$AGB stars show both short and long$-$term
variability in the absorption and emission line strengths and profiles.
P$-$Cygni profiles detected in these stars are also expected to vary as the
stellar wind and post$-$AGB mass loss rates may show variations as the star
evolves.  The line identifications are presented in Tables 2 to 5 and are based
on the Moore multiplet table (1945) and the linelists of Parthasarathy et al.
(2000b),  Klochkova et al. (2002) and Sarkar et al. (2005). Unidentified lines
are denoted by ``UN''. Night sky emission lines denoted by ``atmos.'' were
identified from  Osterbrock et al. (1996). The laboratory wavelengths, log~(gf)
values, and excitation potentials ($\chi$) have been extracted from the Kurucz
(CD$-$ROM 23) linelist  GFALL (Moore 1945). Ivan Hubeny and Thierry Lanz
have compiled the Kurucz  linelists with improved oscillator strengths from the 
NIST Atomic Spectra Database. For wavelengths below 7500 \AA~ we have used their
data which may be retrieved from
http://nova.astro.umd.edu/Synspec43/synspec$-$frames$-$data.html. Note also
that in Table 2 the rest wavelengths from Hobbs et al. (2008) are given for the
diffuse interstellar bands (DIBs) identified in I22023.

\setcounter{table}{1}
\begin{table*}
\centering
 \begin{minipage}{140mm}
  \caption{Absorption lines in IRAS 22023$+$5249$^{a}$}
\begin{tabular}{|c|c|c|c|c|c|c|c|}
\hline
${\rm \lambda}_{\rm obs.}$ & ${\rm \lambda}_{\rm lab.}$ & Ident. & W$_{\rm \lambda}$ &
log (gf) & $\chi$ & $|\Delta {\rm \lambda}|$ & V$_{\rm r}$ \\
   (\AA~)            &   (\AA~)             &        &   (\AA~)       &
         &  (eV)  &      (\AA~)      & km s$^{-1}$   \\
\hline 
\hline
4314.874 & 4317.139       & OII(2)       & 0.1609 & $-$0.386 & 22.95$-$25.82 & 2.265 & $-$142.53\\
4317.277 & 4319.630       & OII(2)       & 0.2197 & $-$0.380 & 22.96$-$25.83 & 2.353 & $-$148.55\\
4343.245 & 4345.560       & OII(2)       & 0.1448 & $-$0.346 & 22.96$-$25.81 & 2.315 & $-$144.95\\
4345.141 & 4347.413       & OII(16)      & 0.0480 &    0.024 & 25.64$-$28.49 & 2.272 & $-$141.91\\
4347.137 & 4349.426       & OII(2)       & 0.2489 &    0.060 & 22.98$-$25.83 & 2.289 & $-$143.01\\ 
4348.903 & 4351.260       & OII(16)      & 0.0483 &    0.227 & 25.64$-$28.49 & 2.357 & $-$147.64\\
4364.653 & 4366.895       & OII(2)       & 0.1130 & $-$0.348 & 22.98$-$25.82 & 2.242 & $-$139.15\\
4385.301 & 4387.929       & HeI(51)      & 0.3722 & $-$0.883 & 21.20$-$24.03 & 2.628 & $-$164.80\\
4412.525 & 4414.899$^{b}$ & OII(5)       &        &    0.172 & 23.42$-$26.23 &       & blend  \\
4414.702 & 4416.975$^{c}$ & OII(5)       & 0.1319 & $-$0.077 & 23.40$-$26.21 & 2.384 & blend  \\
4435.158 & 4437.551       & HeI(50)      & 0.0786 & $-$2.034 & 21.20$-$23.99 & 2.393 & $-$146.91\\
4478.638 & 4479.885       & AlIII(8)     & 0.0524 &    0.900 & 20.77$-$23.53 &       &  \\
         & $+$ 4479.971   & AlIII(8)     &        &    1.020 & 20.77$-$23.53 &       &  \\
         & $+$ 4481.126   & MgII(4)      &        &    0.740 &  8.86$-$11.62 & 2.488 & $-$151.70\\
4550.109 & 4552.622       & SiIII(2)     & 0.3600 &    0.181 & 19.00$-$21.72 & 2.513 & $-$150.73\\
4565.371 & 4567.840       & SiIII(2)     & 0.3225 & $-$0.039 & 19.00$-$21.72 & 2.469 & $-$147.28\\
4572.278 & 4574.757       & SiIII(2)     & 0.1638 & $-$0.509 & 19.00$-$21.71 & 2.479 & $-$147.70\\
4588.513 & 4590.974       & OII(15)      & 0.1206 &    0.350 & 25.64$-$28.34 & 2.461 & $-$145.95\\
4593.736 & 4596.177       & OII(15)      & 0.1507 &    0.200 & 25.64$-$28.34 & 2.441 & $-$144.46\\
4627.829 & 4630.539       & NII(5)       & 0.0983 &    0.094 & 18.47$-$21.14 & 2.710 & $-$160.70\\
4636.365 & 4638.856       & OII(1)       & 0.1682 & $-$0.332 & 22.95$-$25.62 & 2.491 & $-$146.23\\
4639.279 & 4641.810       & OII(1)       & 0.3179 &    0.055 & 22.96$-$25.63 & 2.531 & $-$148.71\\
4644.997 & 4647.418       & CIII(1)      & 0.1181 &    0.070 & 29.51$-$32.18 & 2.421 & $-$141.41\\
4646.569 & 4649.135       & OII(1)       & 0.3387 &    0.308 & 22.98$-$25.65 & 2.566 & $-$150.71\\ 
4648.297 & 4650.16        & CIII(1)      & 0.2709 &          &               &       & blend  \\    
         & $+$ 4650.841   & OII(1)       &        &          &               &       &        \\
         & $+$ 4651.35    & CIII(1)      &        &          &               &       &        \\ 
4659.126 & 4661.632       & OII(1)       & 0.2431 & $-$0.278 & 22.96$-$25.62 & 2.506 & $-$146.40\\
4671.265 & 4673.733       & OII(1)       & 0.0644 & $-$1.090 & 22.96$-$25.61 & 2.468 & $-$143.55\\
4673.706 & 4676.235       & OII(1)       & 0.1598 & $-$0.394 & 22.98$-$25.63 & 2.529 & $-$147.38\\
4693.804 & 4696.353       & OII(1)       & 0.0346 & $-$1.380 & 22.98$-$25.62 & 2.558 & $-$148.53\\
4696.824 & 4699.218       & OII(25)      & 0.0944 &    0.270 & 26.21$-$28.84 & 2.394 & $-$137.96\\
4702.873 & 4705.346       & OII(25)      & 0.0621 &    0.477 & 26.23$-$28.86 & 2.473 & $-$142.80\\
4817.027 & 4819.712       & SiIII(9)     & 0.0556 &    0.750 & 25.96$-$28.53 & 2.685 & $-$152.26\\
4826.346 & 4828.951       & SiIII(9)     & 0.0344 &    1.090 & 25.97$-$28.53 & 2.605 & $-$146.97\\
4904.197 & 4906.830$^{d}$ & OII(28)      &        & $-$0.161 & 26.29$-$28.81 & 2.633 & $-$146.11\\  
4918.616 &     4920.35       & HeI(49)  & 0.4414 &          & 21.13$-$23.64 &       & blend  \\  
         & $+$ 4921.931      & HeI(48)  &        & $-$0.435 & 21.20$-$23.72 &       &        \\  
4921.878 &     4924.529      & OII(28)  & 0.0851 &    0.074 & 26.29$-$28.80 & 2.651 & $-$146.63\\  

4940.268 &     4943.005      & OII(33)  & 0.0825 &    0.239 & 26.54$-$29.05 & 2.737 & $-$151.24 \\
5000.141 &     5002.703      & NII(4)   & 0.0328 & $-$1.022 & 18.45$-$20.92 & 2.562 & $-$138.77\\
5002.414 &     5005.150      & NII(19)  & 0.0336 &    0.594 & 20.65$-$23.13 & 2.736 & $-$149.12\\
5007.895 &     5010.621      & NII(4)   & 0.0851 & $-$0.607 & 18.45$-$20.92 & 2.726 & $-$148.34\\
5042.253 &     5044.8        & CII(35)  & 0.1148 &          &               &       & blend \\
         & $+$ 5045.098      & NII(4)   &        &          &               &       &       \\ 
5044.65  &     5047.2        & CII(35)  & 0.0825 &          &               &       & blend \\ 
         & $+$ 5047.736      & HeI(47)  &        &          &               &       &       \\
5130.277 &     5132.947      & CII(16)  & 0.0376 & $-$0.211 & 20.69$-$23.10 &       & blend \\
         & $+$ 5133.282      & CII(16)  &        & $-$0.178 & 20.69$-$23.10 &       &       \\
5140.722 &     5143.495      & CII(16)  & 0.0640 & $-$0.212 & 20.69$-$23.10 &       & blend \\
5142.407 &     5145.165      & CII(16)  & 0.0523 &    0.189 & 20.70$-$23.10 &       & blend \\
5148.32  &     5151.085      & CII(16)  & 0.0446 & $-$0.179 & 20.70$-$23.10 & 2.765 & $-$146.16\\
5216.486 &                   & UN       & 0.0471 &          &               &       &        \\
5486.97  &     5487.69       & DIB      & 0.0343 &          &               & 0.460 & $-$10.28 \\ 
5663.356 &     5666.630      & NII(3)   & 0.1400 & $-$0.045 & 18.45$-$20.64 & 3.274 & $-$158.46\\
5672.88  &     5676.02       & NII(3)   & 0.1433 & $-$0.367 & 18.45$-$20.63 & 3.14  & $-$151.09\\
5676.262 &     5679.56       & NII(3)   & 0.2223 &    0.250 & 18.47$-$20.65 & 3.298 & $-$159.33\\
5682.922 &     5686.21       & NII(3)   & 0.0382 & $-$0.549 & 18.45$-$20.63 & 3.288 & $-$158.60\\
\hline
\end{tabular}
\end{minipage}
\end{table*}

\setcounter{table}{1} 
\begin{table*} 
\centering 
\begin{minipage}{140mm}
\caption{Absorption lines in IRAS 22023$+$5249$^{a}$. contd..}
\begin{tabular}{|c|c|c|c|c|c|c|c|} \hline ${\rm \lambda}_{\rm obs.}$ & ${\rm
\lambda}_{\rm lab.}$ & Ident. & W$_{\rm \lambda}$ & log (gf) & $\chi$ & $|\Delta
{\rm \lambda}|$ & V$_{\rm r}$ \\ (\AA~)            &   (\AA~)            
&        &   (\AA~)       & &  (eV)  &      (\AA~)      & km s$^{-1}$   \\
\hline  \hline
5693.475 &     5695.920      & CIII(2)  & 0.1418 &    0.017 & 32.08$-$34.26 &       & \\
         & $+$ 5696.604      & AlIII(2) &        &    0.230 & 15.63$-$17.81 & 3.129 & $-$149.91\\
5707.504 &     5710.770      & NII(3)   & 0.0552 & $-$0.518 & 18.47$-$20.64 & 3.266 & $-$156.70\\
5719.637 &     5722.730      & AlIII(2) & 0.0786 & $-$0.070 & 15.63$-$17.80 & 3.093 & $-$147.27\\
5736.688 &     5739.734      & SiIII(4) & 0.2575 & $-$0.160 & 19.71$-$21.87 & 3.046 & $-$144.34\\
5779.728 &     5780.480      & DIB      & 0.3405 &          &               & 0.682 & $-$20.52 \\
5796.444 &     5797.060      & DIB      & 0.0440 &          &               & 0.586 & $-$15.46 \\
6139.673 &     6143.063      & NeI(1)   & 0.0282 & $-$0.350 & 16.61$-$18.62 & 3.39  & $-$150.68\\
6195.200 &     6195.980      & DIB      & 0.0756 &          &               & 0.790 & $-$23.38 \\
6202.191 &     6203.050      & DIB      & 0.0966 &          &               & 0.869 & $-$27.16 \\
6269.073 &     6269.850      & DIB      & 0.0691 &          &               & 0.777 & $-$22.50  \\
6282.734 &     6283.840$^{e}$ & DIB      & 0.7319 &          &               & 1.126 & $-$38.89 \\
6398.61  &     6402.246      & NeI (1)  & 0.0543 & 0.360    & 16.61$-$18.54 & 3.636 & $-$155.48\\
6612.857 &     6613.620      & DIB      & 0.1268 &          &               & 0.773 & $-$20.19 \\ 
6637.522 &     6641.031      & OII(4)   & 0.0238 & $-$0.884 & 23.40$-$25.27 & 3.509 & $-$143.64\\
6717.943 &     6721.388      & OII(4)   & 0.0963 & $-$0.610 & 23.43$-$25.27 & 3.445 & $-$138.89\\
7766.623 &     7771.944      & OI(1)    & 0.576  &    0.320 &  9.14$-$10.74 &       & blend  \\
7769.416 &     7774.166      & OI(1)    & 0.434  &    0.170 &  9.14$-$10.74 &       & blend  \\
         & $+$ 7775.388      & OI(1)    &        & $-$0.050 &  9.14$-$10.74 &       &        \\
8592.524 &                   & UN       & 0.265  &          &               &       &        \\
8646.868 &     8648.280      & DIB      & 0.466  &          &               & 1.412 & $-$34.11 \\
\hline
\end{tabular}
\indent \parbox{12cm}{$^{a}$ Note that the rest wavelengths from Hobbs et
al. (2008) are given for the DIBs. $^{b}$OII(5) 4414.899\AA~ is blended with 
FeII(32) \AA~ emission. $^{c}$OII(5) 4416.975\AA~ is blended with the absorption
component of FeIII(114) 449.596 P$-$Cygni profile. $^{d}$OII(28) 4906.88 \AA~ is
blended  with [FeII](20F) 4905.35 \AA~ emission. $^{e}$DIB 6283.84\AA~ is
blended with telluric absorption lines in this region.}
\end{minipage}
\end{table*}

\setcounter{table}{2} 
\begin{table*}
\centering
 \begin{minipage}{140mm}
  \caption{Emission lines in IRAS 22023+5249}
\begin{tabular}{|c|c|c|c|c|c|c|c|}
\hline
${\rm \lambda}_{\rm obs.}$ & ${\rm \lambda}_{\rm lab.}$ & Ident. & W$_{\rm \lambda}$ &
log (gf) & $\chi$ & $|\Delta {\rm \lambda}|$ & V$_{\rm r}$ \\
   (\AA~)            &   (\AA~)             &        &   (\AA~)       &
         &  (eV)  &      (\AA~)      & km s$^{-1}$   \\
\hline 
\hline
4411.134 & 4413.601$^{a}$ & FeII(32)       & 0.0600 & $-$3.870 &  2.67$-$5.48  &       & blend   \\
4811.856 & 4814.55        & [FeII](20F)    & 0.0409 &          &               & 2.694 & $-$153.00\\ 
4902.59  & 4905.35$^{b}$  & [FeII](20F)    & 0.0287 &          &               &       & blend \\ 
5038.313 & 5041.024       & SiII(5)        & 0.1095 &    0.291 & 10.06$-$12.52 & 2.711 & $-$146.47\\ 
5053.306 & 5055.984       & SiII(5)        & 0.2758 &    0.593 & 10.07$-$12.52 &       & blend  \\
         & $+$ 5056.317   & SiII(5)        &        & $-$0.359 & 10.07$-$12.52 &       &       \\
5088.76  &                &  UN            & 0.0244 &          &               &       &       \\
5155.834 & 5158.81        & [FeII](19F)    & 0.0985 &          &               &       & blend \\ 
5191.418 & 5193.909       & FeIII(5)       & 0.0594 & $-$2.852 &  8.65$-$11.04 &       & blend  \\
         & $+$ 5194.384   & FeIII(5)       &        &          &  8.65$-$11.04 &       &        \\
5195.026 & 5197.929       & FeI(1091)      & 0.1135 & $-$1.640 &  4.30$-$6.68  & 2.903 & $-$152.68\\ 
5197.427 &                & UN             & 0.084  &          &               &       &        \\
5232.982 &                & UN             & 0.0412 &          &               &       &        \\
5240.659 & 5243.306       & FeIII(113)     & 0.0997 &    0.405 & 18.26$-$20.62 &       &  blend \\
         & $+$ 5243.773   & FeI(1089)      &        & $-$1.150 &  4.25$-$6.62  &      &         \\
5258.712 & 5261.61        & [FeII](19F)    & 0.0520 &          &               & 2.898 & $-$150.36\\
5270.327 & 5273.38        & [FeII](18F)    & 0.0481 &          &               & 3.053 & $-$158.81\\
5273.648 &                & UN             & 0.0487 &          &               &       &        \\
5279.665 &                & UN             & 0.0756 &          &               &       &        \\
5282.071 &                & UN             & 0.0277 &          &               &       &        \\
5286.85  &                & UN             & 0.0632 &          &               &       &        \\ 
5296.105 & 5299.044       & OI(26)         & 0.1163 & $-$2.140 & 10.98$-$13.32 &       & blend  \\
5297.325 &                & UN             & 0.0366 &          &               &       &        \\
5299.853 &                & UN             & 0.0368 &          &               &       &        \\
5751.43  & 5754.8         & [NII](3F)      & 0.0436 &          &               &       & weak   \\
5830.998 & 5834.06        & FeII(165)      & 0.0705 & $-$3.738 &  5.57$-$7.69  & 3.062 & $-$142.58\\
5917.114 & 5920.124       & FeIII(115)     & 0.1332 & $-$0.034 & 18.78$-$20.87 & 3.01  & $-$137.66\\
5926.609 & 5929.685       & FeIII(114)     & 0.0964 &    0.351 & 18.50$-$20.59 & 3.076 & $-$140.75\\
5950.424 & 5953.613       & FeIII(115)     & 0.0989 &    0.186 & 18.78$-$20.86 & 3.226 & $-$147.69\\
5954.336 & 5957.559       & SiII(4)        & 0.1449 & $-$0.301 & 10.06$-$12.14 & 3.223 & $-$147.43\\
5975.851 & 5978.930       & SiII(4)        & 0.4134 &    0.004 & 10.07$-$12.14 & 3.079 & $-$139.62\\
5996.369 & 5999.70        & AlII(93)       & 0.1843 &          & 15.52$-$17.57 &       & blend   \\
         & $+$ 5999.83    & AlII(93)       &        &          & 15.52$-$17.57 &       &         \\
6029.382 & 6032.67        & FeI(1082)      & 0.2320 &          &  4.20$-$6.25  & 3.288 & $-$148.64\\
6043.111      & 6046.233       & OI(22)       & 0.0968 & $-$1.895 & 10.98$-$13.03 &       & blend   \\
              & $+$ 6046.438   &              &        & $-$1.675 & 10.98$-$13.03 &       &         \\
6092.139      & 6095.290       & CII(24)      & 0.0131 & $-$0.029 & 22.55$-$24.58 & 3.151 & $-$140.22\\ 
6095.324      & 6098.510       & CII(24)      & 0.0344 &    0.226 & 22.56$-$24.59 & 3.186 & $-$141.86\\
6146.784      & 6150.10        & FeII(46)     & 0.0292 &          & 3.21$-$5.21   &       & blend \\ 
6148.099      &                & UN           & 0.0828 &          &               &       &         \\
6296.769      & 6300.23        & [OI](1F)     & 0.2832 &          &               & 3.461 & $-$149.93\\
6337.276      & 6340.58        & NII(46)      & 0.0479 & $-$0.192 & 23.23$-$25.18 & 3.304 & $-$141.46\\
6343.686      & 6346.86        & NII(46)      & 0.5132 & $-$0.901 & 23.22$-$25.18 &       & blend   \\
              & $+$ 6347.109   & SiII(2)      &        &    0.297 & 8.12$-$10.07  &       &         \\
6353.584      & 6357.0         & NII(46)      & 0.0511 &          & 23.23$-$25.18 & 3.416 & $-$146.34\\
6360.279      & 6363.88        & [OI](1F)     & 0.0677 &          &               & 3.601 & $-$154.89\\
6367.944      & 6371.371       & SiII(2)      & 0.1989 & $-$0.003 & 8.12$-$10.06  & 3.427 & $-$146.49\\
6458.428      &                & UN           & 0.1059 &          &               &       &         \\ 
6544.492      & 6548.1         & [NII](1F)    & 2.817  &          &               & 3.608 & $-$150.43 \\
6579.838      & 6583.6$^{c}$   & [NII](1F)    & 8.534  &          &               &       & blend   \\
6712.772      & 6717.0         & [SII](2F)    & 0.5017 &          &               & 4.228 & $-$173.96\\
6727.154      & 6731.3         & [SII](2F)    & 1.007  &          &               & 4.146 & $-$169.91\\
6848.05       & 6851.634       & FeI(34)      & 0.1321 & $-$5.320 &  1.61$-$3.41  & 3.584 & $-$142.06 \\
6998.28        & 7001.93        & OI(21)      & 0.0786 &          & 10.94$-$12.70 &       & blend     \\
               & $+$ 7002.22    & OI(21)      &        &          & 10.94$-$12.70 &       &           \\
               & 7231.330$^{d}$ & CII(3)      &        &    0.043 & 16.32$-$18.03 &       & blend     \\
               & 7236.420$^{d}$ & CII(3)      &        &    0.299 & 16.32$-$18.03 &       & blend     \\
               & 7316.282$^{d}$ & atmos.      &        &          &               &       &           \\ 
7373.821       &                & UN          & 0.1893 &          &               &       &           \\
\hline
\end{tabular}
\end{minipage}
\end{table*}

\setcounter{table}{2} 
\begin{table*}
\centering
 \begin{minipage}{140mm}
  \caption{Emission lines in IRAS 22023+5249. contd.}
\begin{tabular}{|c|c|c|c|c|c|c|c|}
\hline
${\rm \lambda}_{\rm obs.}$ & ${\rm \lambda}_{\rm lab.}$ & Ident. & W$_{\rm \lambda}$ &
log (gf) & $\chi$ & $|\Delta {\rm \lambda}|$ & V$_{\rm r}$ \\
   (\AA~)            &   (\AA~)             &        &   (\AA~)       &
         &  (eV)  &      (\AA~)      & km s$^{-1}$   \\
\hline 
\hline
7458.676       &                & UN          & 0.1119 &          &               &       &           \\
7462.426       &                & UN          & 0.1229 &          &               &       &           \\ 
7464.227       &                & UN          & 0.122  &          &               &       &           \\
7892.292       &                & UN          & 0.134  &          &               &       &           \\
7993.283       & 7993.332       & atmos.      & 0.0534 &          &               &       &           \\
8218.700       & 8223.128$^{d}$ & NI(2)       &        & $-$0.390 & 10.32$-$11.83 &       & blend     \\
8237.897       & 8242.389$^{d}$ & NI(2)       &        & $-$0.380 & 10.33$-$11.83 &       & blend     \\ 
8283.265       &                & UN          &        &          &               &       & blend     \\
8441.864       & 8446.359       & OI(4)       & 3.097  &    0.170 &  9.51$-$10.98 &       & blend     \\ 
               & $+$ 8446.758   & OI(4)       &        & $-$0.050 &  9.51$-$10.98 &       &           \\ 
8745.796$^{d}$ &                & UN          &        &          &               &       &           \\
8857.744$^{d}$ &                & UN          &        &          &               &       &           \\
\hline
\end{tabular}
\indent \parbox{12cm}{$^{a}$FeII(32) 4413.601\AA~ is blended with OII(5)
4414.899\AA~ absorption feature. $^{b}$[FeII](20F) 4905.35\AA~ is blended with
OII(28) 4906.88\AA~ absorption feature. $^{c}$[NII](1F) 6583.6\AA~ is blended
with the emission component  of CII(2) 6582.88\AA~ P$-$Cygni profile.
$^{d}$These emission lines are weak and are blended with the atmospheric
absorption lines in this region.} 
\end{minipage}
\end{table*}

\subsection{Description of the spectrum}

The absorption and emission lines in the spectrum of I22023 are similar to 
those detected in other hot post$-$AGB stars (Parthasarathy et al. 2000b, 
Klochkova et al. 2002 and Sarkar et al. 2005). In addition to the O~I triplet,
absorption lines  of He~I, C~II, C~III, N~II, O~II, Ne~I, Mg~II, Al~III, Si~III
were identified.  Emission lines of C~II, N~II, O~I, [O~I], Al~III, Si~II, Fe~I,
Fe~II, [Fe~II] and Fe~III were also identified in the spectrum of the star. The
NaI D1 and D2 lines show a complex structure.  The presence of low excitation
nebular lines  of [N~II] and [S~II] and the absence of [O~III] 5007\AA~ suggest
that photoionization has just started, although shock excitation may also be
playing a role as indicated by the 40\% thermal (shock) excitation to the
observed H$_{2}$ emission (Kelly \& Hrivnak 2005). Balmer lines of H$_{\alpha}$,
H$_{\beta}$ and H$_{\gamma}$ show P$-$Cygni profiles indicating ongoing
post$-$AGB mass loss. Some He~I, C~II and Fe~III lines were also found to have
P$-$Cygni profiles. 

\subsection{Radial velocity}

Heliocentric radial velocities have been derived from  wavelength shifts of the
well defined absorption and emission  lines (Tables 2, 3, 4, and 5). The mean
heliocentric radial velocities from the absorption and emission lines (Tables 2
and 3) are $-$148.31 $\pm$ 0.60 kms$^{-1}$  and $-$144.13 $\pm$ 0.72 kms$^{-1}$,
respectively.  Radial velocity measurements of the forbidden lines have been
excluded in estimating the mean.  The quoted errors refer to the probable errors
of estimation. Figure 1 shows the radial velocity trends with respect to the
equivalent widths (W$_{\lambda}$) and lower excitation potentials (LEP) of the
absorption and emission lines, respectively.

The mean heliocentric radial velocity of the [N~II], [O~I] and  [Fe~II] lines is
$-$152.90 $\pm$ 0.96 kms$^{-1}$. [S~II] 6717.0\AA~ and 6731.3\AA~ lines in
I22023 have a markedly different heliocentric velocity  corresponding to
$-$171.93 $\pm$ 1.36 kms$^{-1}$. The different radial velocities argue for a
non-spherical nebula.

\begin{figure*}
\epsfig{figure=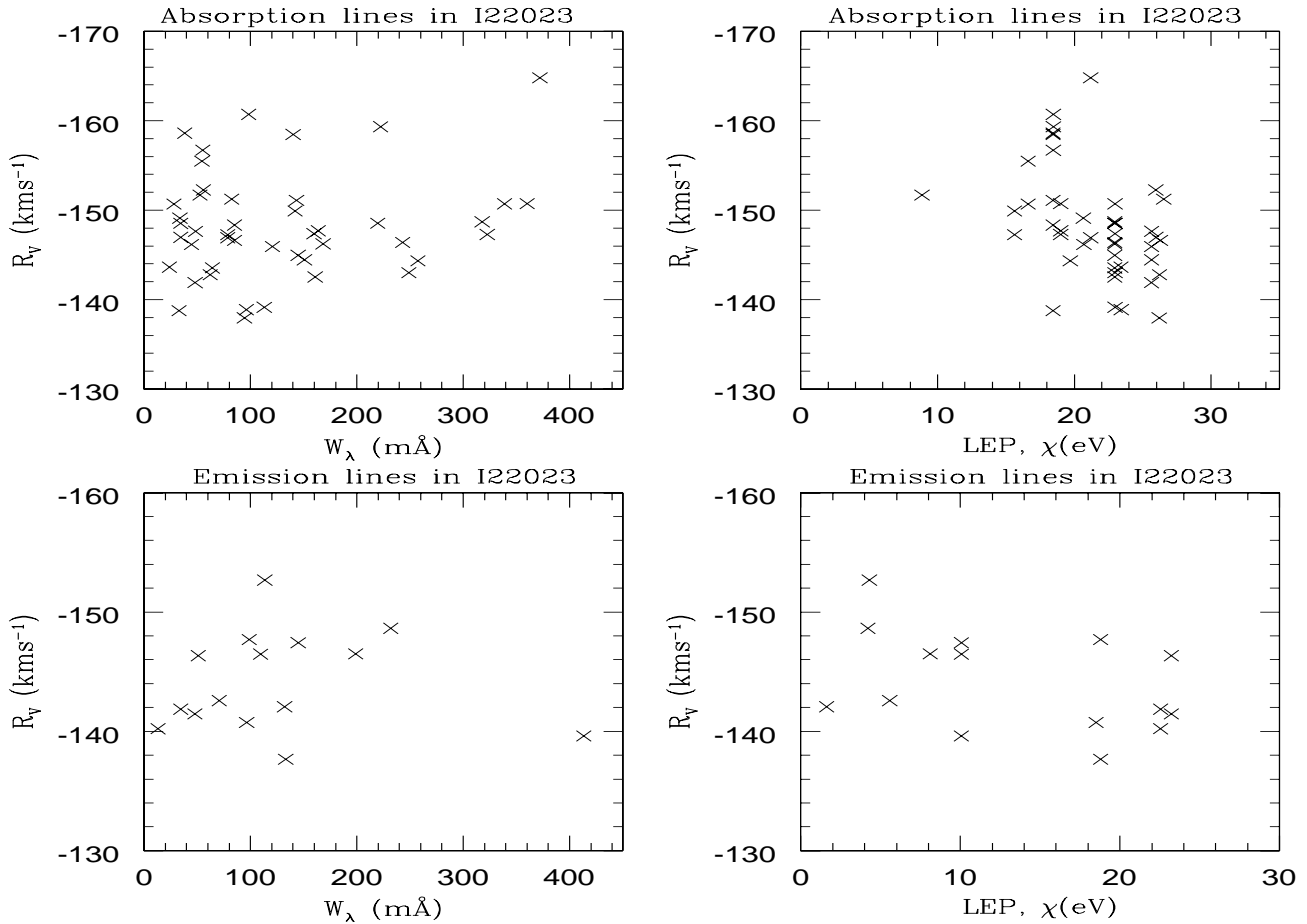, width=18cm, height=13cm}
\caption{Radial velocity trends of the absorption and emission
lines in IRAS22023+5249. Radial velocity measurements of the forbidden
lines have not been plotted.}
\end{figure*}

\subsection{Wind velocities and mass loss rate from the P$-$Cygni profiles}

The Balmer lines, H$_{\alpha}$, H$_{\beta}$, H$_{\gamma}$, a few He~I, C~II and
Fe~III lines in I22023 show P$-$Cygni behaviour (see the appendix and Table 4)
indicating ongoing mass$-$loss. Wind velocities were estimated from the blue
absorption edges of the well defined and unblended P$-$Cygni profiles (Table 4).
The wind velocities of the Fe~III(5) lines are markedly different from those of
the other species. However, the wind velocities do not show any obvious trend
with the lower excitation potentials or ionization potentials of the species.
The mean wind velocity from the Balmer, He~I and C~II lines is 187.48 kms$^{-1}$
and that from the Fe~III(5) lines is 149.26 kms$^{-1}$. This difference in
velocities indicates a deviation from spherical symmetry -- possibly a bipolar 
morphology and the presence of a dense equatorial torus (e.g., Welch et al.
1999; Sahai et al. 2005). Only very high-spatial resolution (FWHM$<$0.15")
images (e.g., using the Hubble Space Telescope) may reveal the true morphology
of the object.

The Balmer lines H$_{\alpha}$, H$_{\beta}$, and H$_{\gamma}$ in I22023 are shown
in Figure 2 where the velocities are in the heliocentric frame. A more detailed
modelling of these lines to derive the mass loss rate is out of the scope of
this paper. The equivalent widths of the H$_{\alpha}$ emission components are
related to the mass loss rates in OB stars (Leitherer 1988). The  H$_{\alpha}$
emission component in I22023 has an equivalent width  (W$_{\lambda}$=37.8\AA)
comparable to that of the O8I star, HD152408 (W$_{\lambda}$=34.7\AA). From this,
we estimate a mass loss rate of  1.23$\times$10$^{-5}$ M$_{\odot}$yr$^{-1}$
(Leitherer 1988). However, the Leitherer (1988)'s method is valid for massive
stars and it may be not applicable for hot post-AGB supergiants. Therefore the
mass loss rate estimated for I22023 may be not correct and hence it should be
used with caution. Also, in I22023, the large equivalent width of the 
H$_{\alpha}$ emission component may be due to a large amount of gas  in its
circumstellar envelope or to the presence of a possible bipolar envelope (Volk
et al. 2004) and may not be directly related to the mass loss rate.

\begin{table*}
\centering
 \begin{minipage}{140mm}
  \caption{P$-$Cygni lines in IRAS22023+5249. Equivalent widths of the absorption 
and emission components of the P$-$Cygni profiles are given. Wind velocities
are estimated from the blue absorption edges of the P$-$Cygni profiles.}
\begin{tabular}{|c|c|c|c|c|c|c|}
\hline
${\rm \lambda}_{\rm lab}$ & Ident. & W$_{\rm \lambda}$ (absorption) & W$_{\rm \lambda}$ (emission) & log (gf) &
$\chi$ & Wind Velocity\\
       (\AA~)       &        &  (\AA~)                    & (\AA~)                   &          &
(eV) & km s$^{-1}$\\
\hline \hline
4340.462       & H$_{\gamma}$ & 0.4908 & 1.915  & $-$0.447 & 10.19$-$13.04 & $-$187.81\\
4419.596$^{a}$ & FeIII(4)     & 0.0991 & 0.0987 & $-$2.218 &  8.23$-$11.04 & blend\\
4431.019       & FeIII(4)     & 0.0429 & 0.0506 & $-$2.572 &  8.24$-$11.04 & weak\\
4471.477       & HeI(14)      & 0.3398 & 0.0828 &          & 20.95$-$23.72 & blend\\
$+$4471.682$^{c}$   & HeI(14)      &        &        & $-$0.898 & 20.95$-$23.72 & \\ 
4861.323$^{b}$ & H$_{\beta}$  & 0.1235 & 6.779  & $-$0.020 & 10.19$-$12.74 & $-$192.63\\
5015.678       & HeI(4)       & 0.3955 & 0.5047 & $-$0.820 & 20.60$-$23.07 & $-$185.39\\
5073.903       & FeIII(5)     & 0.1361 & 0.0426 & $-$2.557 &  8.65$-$11.09 & weak\\
5086.701       & FeIII(5)     & 0.0319 & 0.0338 & $-$2.590 &  8.65$-$11.09 & weak\\
5127.387       & FeIII(5)     & 0.1048 & 0.1684 & $-$2.218 &  8.65$-$11.07 & $-$149.39\\
5156.111       & FeIII(5)     & 0.1105 & 0.1964 & $-$2.018 &  8.64$-$11.04 & $-$149.14\\
5875.618       & HeI(11)      & 0.5500 & 1.8790 &          & 20.87$-$22.97 & blend\\
$+$ 5875.650$^{c}$   & HeI(11)      &        &        &          & 20.87$-$22.97 & \\
$+$ 5875.989$^{c}$   & HeI(11)      &        &        &          & 20.87$-$22.97 & \\
6562.797$^{b}$ & H$_{\alpha}$ & 0.0443 & 37.86  & 0.710    & 10.19$-$12.08 & $-$185.36\\
6578.050       & CII(2)       & 0.1977 & 0.2779 & $-$0.026 & 14.44$-$16.32 & $-$191.55\\
6582.880$^{d}$ & CII(2)       &        &        & $-$0.327 & 14.43$-$16.32 & blend\\
6678.154       & HeI(46)      & 0.6693 & 0.8538 &    0.329 & 21.20$-$23.06 & $-$182.16\\
7065.176       & HeI(10)      & 0.3406 & 1.3600 & $-$0.460 & 20.95$-$22.70 & blend\\
$+$ 7065.707$^{c}$   & HeI(10)      &        &        & $-$1.160 & 20.95$-$22.70 & \\
\hline
\end{tabular}
\indent \parbox{14cm}{$^{a}$The absorption component of FeIII(114) 4419.596\AA~
is blended with OII(5) 4416.975 absorption feature. $^{b}$The emission
components of the H$_{\beta}$  and H$_{\alpha}$ profiles have broad wings.
Gaussian fits to the absorption and emission  components of these profiles could
not be obtained. Using IRAF, the equivalent widths of these components were
estimated by subtracting the linear continuum between  the points of interest
and summing the pixels with partial pixels at the ends.  $^{c}$ CII(2)
6582.88\AA~ P$-$Cygni profile is blended with [NII](1F) 6583.6\AA~.} 
\end{minipage}
\end{table*}

\begin{figure*}
\epsfig{figure=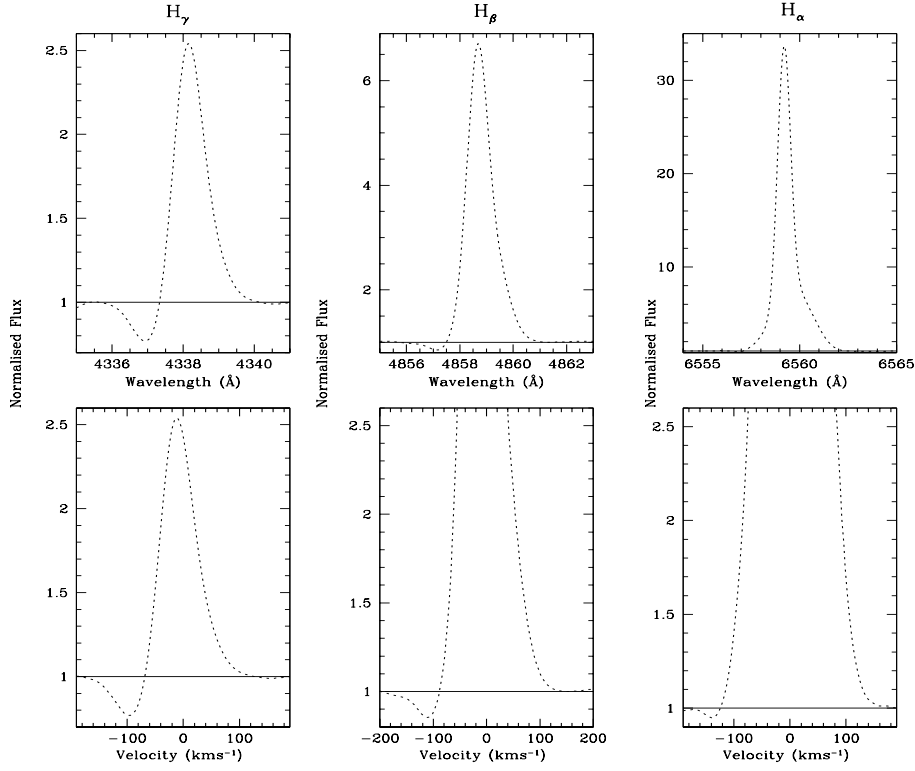,width=13cm,height=11cm}
\caption{Observed Balmer H$_{\gamma}$ H$_{\beta}$, and H$_{\alpha}$ profiles
(dotted) in I22023, showing P-Cygni behaviour. Note that the velocities are in
the heliocentric  frame.}
\end{figure*}

\subsection{Diffuse interstellar bands (DIBs)}

Diffuse interstellar bands (DIBs) at 5487.69 \AA, 5780.48 \AA, 5797.06 \AA,
6195.98 \AA, 6203.05 \AA, 6269.85 \AA, 6283.84 \AA, 6613.620 \AA, etc. (Hobbs et
al. 2008) were identified in the spectrum of the star (see Table 2). DIBs were
also detected in the spectra of several other hot post$-$AGB stars such as IRAS
01005$+$7910 (Klochkova et al. 2002), IRAS 13266$-$5551, and IRAS 17311$-$4924
(Sarkar et al. 2005). Some of us presented a detailed analysis of the most
famous DIBs in the spectrum of I22023 and in the spectra of several other
post-AGB stars (Luna et al. 2008). Luna et al. (2008) found that the DIBs'
strength in post-AGB stars is consistent with the insterstellar extinction
toward these sources. This implies that DIBs are not originated in the
circumstellar shells of post-AGB stars. Thus, we estimate an interstellar
E(B$-$V)=0.67 from the measured equivalent width of 0.3405 \AA~for the 5780
\AA~DIB and the correlation measured by Friedman et al. (2011). As we have
mentioned before, a more detailed analysis of DIBs in post-AGB stars can be
found in the paper by Luna et al. (2008).

\subsection{Na I D$_{2}$ and Na I D$_{1}$ lines}

Six components were identified in the Na I D$_{2}$ and Na I D$_{1}$ lines (see
Figure 3 and Table 5). The velocities of absorption component 1 and emission
component 2 are comparable with the mean heliocentric  radial velocities of the
absorption and emission lines in the star (Sect. 3.2) suggesting that component
1 is of photospheric origin and component 2 arises in an extended envelope
around the central star. Comparing the heliocentric velocities of absorption
components 3, 4, and 5 with those of DIBs observed in the spectrum of the star 
we may infer that these components originate in the interstellar medium.
Component 6 is observed in emission with a velocity very different from the
envelope velocity. This component may arise in a disk or in outflows around the
central star. The velocity of this component is comparable with the expansion
velocity estimated from the nebular lines (Sect. 4.2).

\begin{table*}
\centering
 \begin{minipage}{140mm}
  \caption{Absorption (a) and emission (e) components of Na~I D$_{2}$ (5889.953 \AA~)
and Na~I D$_{1}$ (5895.923 \AA~) lines in the spectrum of IRAS22023+5249 (LSIII +5224).
W$_{\rm \lambda}$ are the equivalent widths of the components and V$_{\rm r}$ are the
respective heliocentric radial velocities.}
\begin{tabular}{|c|c|c|c|c|c|c|c|}
\hline
           & & \multicolumn{6}{c|}{IRAS22023+5249} \\ \cline{3-8}
          & & \multicolumn{3}{c|}{Na~I D$_{2}$} & \multicolumn{3}{c|}{Na~I D$_{1}$} \\ \cline{3-8}
Component & & ${\rm \lambda}_{\rm obs.}$ & W$_{\rm \lambda}$ & V$_{\rm r}$ &
${\rm \lambda}_{\rm obs.}$ & W$_{\rm \lambda}$ & V$_{\rm r}$ \\
          & & (\AA~) & (\AA~) & (km s$^{-1}$) & (\AA~) & (\AA~) & (km s$^{-1}$) \\
\hline
1. & a & 5886.341  & 0.0559 & $-$169.10 & 5892.323 & 0.0529 & $-$168.31\\
2. & e & 5886.811  & 0.1314 & $-$145.17 & 5892.763 & 0.0277 & $-$145.92\\
3. & a & 5888.530  & 0.1395 & $-$57.61  & 5894.500 & 0.1355 & $-$57.54\\
4. & a & 5888.903  & 0.5054 & $-$38.61  & 5894.881 & 0.3684 & $-$38.15\\
5. & a & 5889.500  & 0.6384 & $-$8.20   & 5895.455 & 0.6241 & $-$8.94\\
6. & e & 5889.918  & 0.1221 & $+$13.09  & 5895.902 & 0.1471 & $+$13.8\\
\hline
\end{tabular}
\end{minipage}
\end{table*}

\begin{figure*}
\epsfig{figure=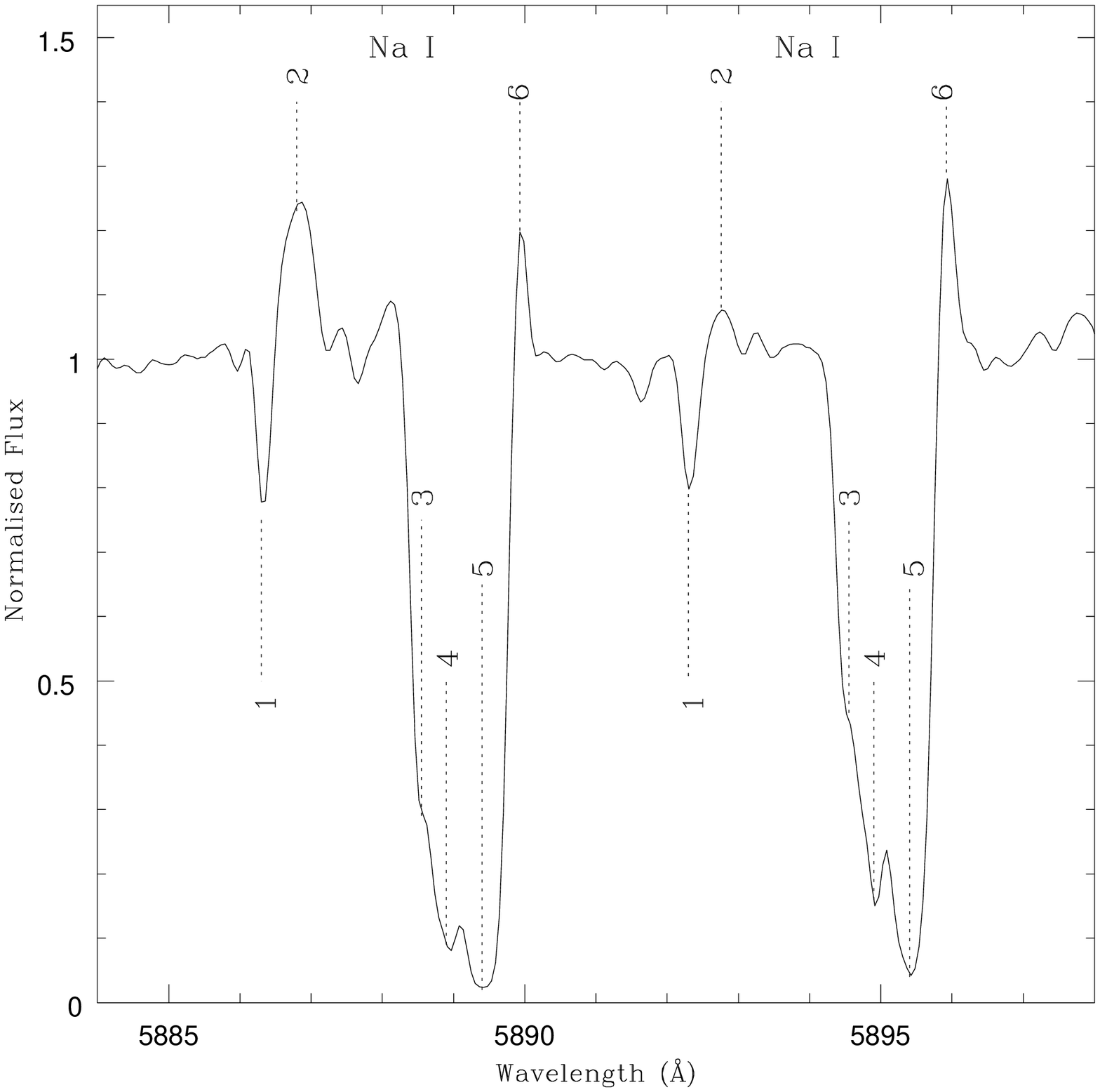,width=13cm,height=11cm}
\caption{Na~I D$_{2}$ and Na~I D$_{1}$ lines in I22023. The various absorption 
and emission components (Sect. 3.5, Table 5) have been labelled.}
\end{figure*}

\section{Analysis of the absorption and emission line spectra} 

\subsection{Atmospheric parameters and abundances from absorption line spectrum}

The detection of He~I and Si~III absorption lines in addition to  the C~III
lines and the absence of He~II absorption lines, indicates a B0 $-$ B1
supergiant spectral type for the central star. A previous comparison of the
UV(IUE) spectrum of I22023 with standard stars suggested that the star was
similar to a B2$-$supergiant (Gauba \& Parthasarathy 2003). Similarly to our
previous analysis of the high-resolution optical spectrum of the hot post$-$AGB
star IRAS 13266$-$5551 (Sarkar et al. 2005), we have used the Kurucz's WIDTH 9
program and the spectrum synthesis code SYNSPEC (Hubeny et al. 1985) together
with solar metallicity  Kurucz (1994) model atmospheres to derive the
atmospheric parameters and elemental abundances of I22023 under the Local
Thermodynamic Equilibrium (LTE) approximation.

The largest number of absorption lines in I22023 are those of O~II and N~II. We
derived the oxygen and nitrogen abundance with the WIDTH 9 program for various
combinations of effective temperature (T$_{eff}$), gravity (log g), and
microturbulence ($\xi_{\rm t}$). We covered 18,000 K $\leq$ T$_{eff}$ $\leq$
25,000 K~and 5 $\leq$ $\xi_{\rm t}$ $\leq$ 10 kms$^{-1}$. From the Kurucz (1994)
model  atmospheres, the log g value was limited to a minimum of 3.0. For each 
combination of these parameters, we then synthesised the spectrum using SYNSPEC.
The best fit to the observed spectrum was obtained for  T$_{eff}$ = 24,000 $\pm$
1000 K, log~g = 3.0 $\pm$ 0.5, $\xi_{\rm t}$ = 7 $\pm$ 1 kms$^{-1}$. Since
strong lines are usually affected by microturbulence, the  use of these lines in
determining the atmospheric parameters of the star may contribute to systematic
errors. Thus, we excluded lines with W$_{\lambda}$ $\ge$ 200 m\AA~in our
estimation of the atmospheric parameters and abundances. Line blends were also
excluded from our analysis. Final abundances for He, C, N, O, Ne, and Si are
summarized in Table 6. The estimated errors in the derived abundances taking
into account  typical variations in the atmospheric parameters are listed in
Table 7.

\begin{table*}
\centering
 \begin{minipage}{140mm}
  \caption{Derived abundances for I22023. The values are for log $\epsilon$(H) = 12.0. 
n refers to the number of lines of each species used for the
abundance determination.  For comparison we have listed the solar abundances
(log $\epsilon$(X)$_{\odot}$)and average abundances for main sequence B$-$stars 
from the Ori OB1 association (Kilian, 1992).}
\begin{tabular}{|c|c|c|c|c|c|c|c|c|c|}
\hline
 & \multicolumn{4}{c|}{IRAS22023+5249} & & \multicolumn{1}{c|}{Main sequence}\\
 & \multicolumn{4}{c|}{($T_{\rm eff}$=24000~K, log $g$=3.0, $\xi_{\rm t}$=7 km s$^{-1}$)}
 & & \multicolumn{1}{c|}{B$-$stars, Ori OB1}\\
X & n & log $\epsilon(X)$ & $\sigma$ & [X/H] & log $\epsilon$(X)$_{\odot}$
& log $\epsilon$(X) \\ \hline \hline

He~I   &  1 & 11.04$^{\dagger}$  & --   & $+$0.11 & 10.93 & 11.04\\ 
C~II   &  4 &  8.58$^{\dagger}$  & --   & $+$0.19 &  8.39 &  8.23\\
N~II   &  7 &  8.36             & 0.33 & $+$0.44 &  7.92 &  7.72\\
O~II   & 18 &  8.90             & 0.42 & $+$0.21 &  8.69 &  8.60\\
Ne~I   &  2 &  9.02             & 0.14 & $+$0.94 &  8.08 &  --  \\
Al~III &  1 & $<$ 6.79$^{\dagger}$ &   &         &  6.47 &  --  \\
Si~III &  3 &  7.43             & 0.59 & $-$0.12 &  7.55 &  --  \\
\hline
\end{tabular}
\indent \parbox{16cm}{The abundances were derived using Kurucz's WIDTH9 program.\\
$^{\dagger}$These values were derived from spectrum synthesis analysis using the SYNSPEC code.}
\end{minipage}
\end{table*}

\begin{table*}
\centering
 \begin{minipage}{140mm}
  \caption{Uncertainties in the derived abundances, $\Delta$log $\epsilon$(X), due to
uncertainties in the model atmospheric parameters}
\begin{tabular}{|c|c|c|c|c|}
\hline
Element & $\Delta$T$_{\rm eff}$ & $\Delta$log~g & $\Delta\xi_{\rm t}$ & $\sigma_{\rm m}$ \\
        & $-$1000K              & $+$0.5        & $+$1 km~s$^{-1}$    & \\ 
\hline \hline
N       & $-$0.14 & $-$0.18 & $-$0.05 & 0.23 \\ 
O       & $+$0.05 & $+$0.08 & $-$0.08 & 0.12 \\          
Ne      & $-$0.15 & $-$0.26 & $-$0.01 & 0.30 \\
Si      & $-$0.04 & $-$0.06 & $-$0.07 & 0.10 \\
\hline
\end{tabular}
\end{minipage}
\end{table*}

\subsubsection{He~I lines}

Several absorption and P$-$Cygni type He~I lines were identified in the spectrum
of I22023. The absorption lines are blends with the exception of the strong
4387.929 \AA~He~I(51) (W$_{\lambda}$=372.2 m\AA) line and the He~I(50) 4437.551
\AA~line (W$_{\lambda}$=78.6 m\AA). For the derived  atmospheric parameters of
the star, and by using spectrum synthesis, we estimated  the helium abundance
from the He~I(50) line (see Table 6). 

\subsubsection{C~II and C~III lines}

Both C~II and C~III absorption lines were detected in the spectrum of the star.
However, the number of these lines in I22023 is low. Furthermore, all the
identified carbon lines are blends with the exception of the C II(16) line at
5151.085 \AA. The carbon abundance was therefore estimated using spectrum
synthesis only.  The region of the C II(16) lines ($\sim$ 5132$-$5151 \AA) was
used for this purpose.

\subsubsection{O II lines and the O~I triplet} 

The largest number of absorption lines in the I22023's spectrum are those of
O~II. The O~II  lines with W$_{\lambda}$ $\ge$ 200 m\AA~may be sensitive to
non-LTE effects and such strong O~II lines were not considered in our oxygen
abundance estimation by using the WIDTH9 program. For the derived atmospheric
parameters and the oxygen abundance of log $\epsilon(O)$=8.90, we could not
obtain a perfect fit to the stronger O II lines with the SYNSPEC code. 

On the other hand, the (total) equivalent width of the O~I triplet in the
spectrum of I22023 is 1.01 \AA. This is comparable to the 0.95 \AA~equivalent
width of the O~I triplet in the spectrum of the B1.5Ia hot post$-$AGB star,
LSII$+$34$^{\circ}$26 (Garc\'ia$-$Lario et al. 1997b; Arkhipova et al. 2001).
The O~I triplet at $\lambda$7773\AA~is known to be sensitive to non-LTE effects.
Indeed, we could not obtain a good fit to the O~I triplet by assuming the oxygen
abundance (log~$\epsilon$(O)=8.90) derived from the O~II lines (Table 6).

\subsubsection{N~II and Ne~I lines}

Several N~II and two Ne~I lines were identified in I22023. The abundances of
these lines were estimated using WIDTH9. Again, the stronger N~II lines with 
W$_{\lambda}$ $\ge$ 200m\AA~were not taken into account in our estimation
(log~$\epsilon$(N)=8.36) of the nitrogen abundance in I22023. 

\subsubsection{Metallic lines}

Only one Mg~II line could be identified in the spectrum of the star. This  line
is blended with Al~III(8). Since the Al~III abundance in I22023  is uncertain
(see below), we did not attempt to estimate the magnesium  abundance from the
blended 4481.126\AA~Mg~II(4) line. Also, four Al~III lines could be identified
in I22023. Three of these lines are clear blends with other atomic species.
Therefore, we estimated the aluminium abundance from the single 5722.730
\AA~Al~III(2) line by using spectrum synthesis and we derived
log~$\epsilon$(Al)=6.79. This abundance from a single line with
W$_{\lambda}$=78.6 m\AA~may be treated as an upper limit. The silicon abundance
([Si/H]=$-$0.12) was derived by using three Si III line and suggests that I22023
may be slightly metal deficient. Finally, it is to be noted here that the iron
abundance could not be estimated since the iron lines in I22023 appear only in
emission or show P$-$Cygni profiles.

\subsubsection{Uncertainties in the abundance determinations}

The standard deviation ($\sigma$) which measures the scatter in the abundances
due to individual lines of a particular species was estimated using WIDTH9
(Table 6). The true error, $\sigma$/$\sqrt{n}$, would be smaller for species
with a greater number of lines (n). Table 7 gives the uncertainties in the
abundances due to typical uncertainties in the model atmospheric parameters
taken for the modelling: $\Delta$T$_{\rm eff}$=$-$1000K, $\Delta$log g=$+$0.5,
and  $\Delta\xi_{\rm t}$=$+$1 kms$^{-1}$. Thus, the formal error (always
$\leq$0.3 dex) in the derived abundances is the quadratic sum of the
uncertainties introduced by typical variations of the atmospheric parameters and
it is given by $\sigma_{\rm m}$ in Table 7.

\subsection{Analysis of the emission line spectrum}

Several permitted and forbidden emission lines were identified in the spectrum
of the star and are listed in Table 3. Nebular parameters and expansion
velocities were determined using the forbidden lines (see below).

\subsubsection{Nebular parameters}

In the absence of a flux calibrated spectrum for I22023, it is not possible to
obtain  the absolute fluxes in the observed emission lines. However,  reliable
emission line flux ratios may be deduced by  combining the observed equivalent
widths (Table 3) with  estimates of the stellar continuum flux distribution in
the regions of the emission lines. The latter were obtained for  the derived
atmospheric parameters of the star (Sect. 4.1) by using the SYNSEPC code and the
Kurucz model atmospheres. The emission line fluxes thus estimated are free from
the effects of both interstellar and circumstellar reddening.

The [S~II] $\lambda6717$/$\lambda$6731 line ratio is an electron density
diagnostic and the [N~II] ($\lambda$6548$+\lambda$6583)/$\lambda$5755 line ratio
is  sensitive to electron temperature. In I22023, the [N~II] 6583.6 \AA~emission
is  blended with the emission component of the C~II(2) 6582.88 \AA~P$-$Cygni
profile.  However, comparing the 6582.88 \AA~C~II(2) profile with the 6578.05
\AA~C~II(2) P$-$Cygni profile (see the appendix), we may conclude that the
contribution of C~II(2) to the [N~II] emission profile is negligible. Using the
NEBULAR analysis package under IRAF, we obtained T$_{\rm e}$ vs. N$_{\rm e}$
contours for the observed [S~II] and [N~II] diagnostic ratios of 0.5 and 166.7,
respectively. From the  intersection of the contours we obtained T$_{\rm
e}$=7,000~K and  N$_{\rm e}$=1.2$\times$10$^{4}$~cm$^{-3}$. The high electron
density is comparable to that measured in the very young and compact PN
Hen3$-$1357 (Parthasarathy et al. 1993; Bobrowsky et al. 1998) which evolved
from the hot post$-$AGB stage into a PN in the 20$-$30 yrs (Parthasarathy et al.
1995).

Unfortunately, we could not derive the nebular C, N, and O abundances, which
could then have been compared with the photospheric abundances to estimate the
amount of material lost by the star during nebular formation and the chemical
composition of the nebula. Such a calculation requires an estimate  of the
H$_{\beta}$ emission line flux. However, H$_{\beta}$ in I22023 shows  a
P$-$Cygni profile and it is not possible to estimate the nebular emission  from
this profile.

\subsubsection{Expansion velocities}

Expansion velocities were estimated from the FWHM of the unblended [O I], [N II]
and [S II] lines using V$_{\rm exp}$=0.50 FWHM (see Table 8). Note that the
[N II](1F) 6583.6 \AA~line is blended with the emission component of C II(2)
6582.88 \AA~P-Cygni profile and has not been used to estimate the expansion
velocity.

This approximation is valid when emission is confined to a thin spherically 
symmetric shell. However, I22023 appears to have an incipient bipolar morphology
in recent ground-based high spatial (FWHM$\sim$0.15") resolution images (Volk et
al. 2004). Furthermore, the observed [O I] 6300.23 \AA~and 6363.88 \AA~line
profiles appear to be asymmetric. Though no obvious line split is observed in
the weak [O I] lines, their asymmetric nature may indicate the presence of a red
and a blue component. This may explain the discrepancy between the expansion
velocities estimated from [O I], [N II], [S II] lines in I22023, the former
being nearly twice that of the latter two species. The mean nebular velocity
based on the [N II] and [S II] lines is 17.5 kms$^{-1}$.

The possible bipolar morphology of this object is not completely established
(Volk et al. 2004). In addition, Cerrigone et al. (2008) studied the radio
continuum emission of this object and they found that ``it is difficult to
interpret the morphology observed in I22023 in the framework of the standard
interacting stellar wind (ISW) model (e.g., Kwok et al. 1978), even invoking a
strong density gradient in the nebula. For this object, a jet would be more
likely the source of the observed morphology".

\begin{table*}
\centering
 \begin{minipage}{140mm}
  \caption{Expansion velocities}
\begin{tabular}{|c|c|c|c|}
\hline
Ident. & ${\rm \lambda}_{\rm lab}$ & FWHM & V$_{\rm exp}$ \\
       &  \AA~                     & \AA~ & km s$^{-1}$\\ 
\hline \hline
6300.23 & [OI](1F)  & 1.334 & 31.76 \\
6363.88 & [OI](1F)  & 1.470 & 34.65 \\
6548.1  & [NII](1F) & 0.844 & 19.34 \\
6717.0  & [SII](2F) & 0.747 & 16.68 \\
6731.3  & [SII](2F) & 0.739 & 16.47 \\
\hline
\end{tabular}
\end{minipage}
\end{table*}

\section{Discussion and conclusions} 

Our analysis of the high-resolution (R$\sim$50,000) optical spectrum of I22023
together with our detailed line identifications confirm that I22023 is a hot
(B-type) O-rich post-AGB star (see below). That I22023 is not a normal
population I B star is also suggested by the large heliocentric radial velocity
of the star ($-$148.31 $\pm$ 0.60 kms$^{-1}$)\footnote{Note that the high radial
velocity suggests that I22023 is an old disk low core mass post-AGB star and
that there are old disk stars whose chemical composition is close to the solar
composition (Furhmann 1998, 2004).}, as measured from the absorption lines
present in its high-resolution optical spectrum. Thus, it is more likely that
I22023 is a post$-$AGB star belonging to the old disk population. The presence
of absorption lines of He~I, C~III, Si~III together with [N~II], [O~I], and
[S~II] emission lines indicate a low-excitation nebula surrounding the early
B$-$type central star. The observed P$-$Cygni profiles in the lines of hydrogen,
He~I, C~II, and Fe~III clearly indicate the presence of a stellar wind with a
significant post-AGB mass-loss rate, providing strong evidence for on-going
post$-$AGB mass loss. 

As a first approximation, using LTE analysis, we estimated T$_{\rm eff}$ =
24,000~K $\pm$ 1000~K, log~g = 3.0 $\pm$ 0.5 and $\xi_{\rm t}$ = 7 $\pm$ 1
kms$^{-1}$. The derived CNO abundances are compared with the average CNO
abundances for main$-$sequence B$-$stars from the Ori OB1 association (Kilian
1992) in Table 6. The CNO abundances indicate that I22023 is an evolved star. We
estimated C/O $\sim$ 0.48, implying that the central star is O-rich and that the
C/O ratio was not altered during the previous AGB phase. Our Si abundance
estimate also suggests that I22023 is only slightly metal-deficient with
[Si/H]=-0.12$\pm$0.10. Our derived abundances can be easily explained if I22023
is the descendant of a low-mass (e.g., below $\sim$1.5 M$_{\odot}$) AGB star of
roughly solar metallicity. Low-mass stars evolve very slowly and are expected to
remain O-rich all the way along the AGB because they experience too few thermal
pulses and the third dredge-up is too inefficient\footnote{Note also that
theoretical models predict a higher efficiency of the dredge-up in low
metallicity atmospheres with respect to those with solar metallicity (e.g.
Lugaro et al. 2003).} to modify the original C/O $<$ 1 ratio (see e.g., Herwig
2005 for a review). However, intermediate-mass (1.5 $<$ M $<$ 3-4 M$_{\odot}$)
AGB stars are converted to carbon and s-process enriched stars (see below) while
massive (M $>$ 3-4 M$_{\odot}$) AGB stars remain also O-rich as a consequence of
the Hot Bottom Burning (HBB) activation and experience a completely different
s-process nucleosynthesis (see e.g., Garc\'{\i}a-Hern\'andez et al. 2006a,
2007a, and references therein). Note that the low-mass interpretation for I22023
would be consistent with the fact that I22023 is an optically bright O-rich
post-AGB star while more massive O-rich post-AGB stars (which experienced HBB in
the AGB) usually are completely obscured in the optical range by their thick
circumstellar envelopes (e.g., Garc\'{\i}a-Hern\'andez et al. 2007b). In short,
attending to the high-resolution spectrum of I22023 only, we may conclude that
I22023 is a low-mass O-rich post-AGB star. 

Even though the IRAS cool (e.g., F, G) and hot (O, B) post$-$AGB stars show
supergiant like spectra, indicating an evolutionary sequence in the transition
region from the AGB to the PN stage, they seem to show fundamental differences
in their chemical compositions. In the high Galactic latitude hot (O-, B-type)
post$-$AGB stars, a severe carbon deficiency (i.e., C/O$<$1) is detected
indicating that they left the AGB before the third dredge$-$up (e.g., Conlon et
al. 1991; McCausland et al. 1992; Moehler \& Heber 1998) has enriched the
stellar surface with the products of the complex nucleosynthesis (e.g., carbon
and heavy s-process elements such as Rb, Zr, Y, Sr, etc.) experienced during the
AGB phase (see e.g., Garc\'{\i}a-Hern\'andez et al. 2007a and references
therein). A similar carbon deficiency is also detected in the hot post$-$AGB
stars in globular clusters  (Moehler et al. 1998; Mooney et al. 2004; Jasniewicz
et al. 2004; Thompson et al. 2006). The only exception to this observational
evidence is the field hot post$-$AGB star, IRAS~01005+7910, which shows an
overabundance of carbon (Klochkova et al. 2002). In contrast, among the IRAS
selected cool (F-, G-type) post$-$AGB stars there is a majority of post-AGB
stars in which a severe carbon deficiency is not detected. The F-, G-type
post$-$AGB stars with the still unidentified 21 micron emission feature show an
overabundance of carbon and heavy s$-$process elements (e.g., Van Winckel \&
Reyniers 2000), confirming that they have experienced s-process nucleosynthesis
and the third dredge$-$up in the previous AGB phase and that they evolved from
intermediate-mass carbon stars.

Interestingly, Cerrigone et al. (2009) found that I22023 is a double-dust
chemistry - i.e., it displays the simultaneous presence of both C-rich and
O-rich dust features - post-AGB star from their analysis of the recent Spitzer
mid-infrared ($\sim$5-40 $\mu$m) spectrum of I22023. The mixed-chemistry is
deduced from the detection of amorphous silicates emission at $\sim$10 microns
together with the classical aromatic infrared bands (AIBs; e.g., at $\sim$6.2,
7.7, 8.6, and 11.3 $\mu$m) usually attributed to carbonaceous compounds. The
origin of the double-dust chemistry is still not very well understood and
several scenarios, including the presence of a binary central system, a late
thermal pulse on the AGB or post-AGB phases, HBB cessation by extreme mass loss,
etc., have been proposed to explain the mixed-chemistry phenomenon observed in
AGB stars (e.g., Garc\'{\i}a-Hern\'andez et al. 2006b), post-AGB stars (e.g.,
Waters et al. 1998; Gielen et al. 2011), and PNe (e.g., Perea-Calder\'on et al.
2009). 

The presence of carbonaceous molecules in an O-rich environment such as that in
I22023 (the central star is also O-rich!) is surprising and puzzling. Cerrigone
et al. (2009) propose that the mixed-chemistry in I22023 (and other hot post-AGB
stars) is due to the presence of a circumbinary disk/torus where O-bearing
molecules would be preserved from the 3$^{rd}$ dredge-up, while the C-bearing
molecules would be formed elsewhere in the outflow. The presence of a binary
companion in I22023 and other hot post-AGB stars cannot be ruled out. Indeed,
the presence of a close companion (at a distance of $\sim$0.4") in the
proto-type hot post-AGB object Hen 3-1357 (the ``Stingray Nebula") is well
known (Bobrowsky et al. 1998). In addition, an spectacular incipient bipolar
morphology is clearly seen in the HST images of Hen 3-1357. The Spitzer
spectrum of Hen 3-1357 (see Perea-Calder\'on et al. 2009) resembles that of
I22023, showing amorphous silicates emission at 10 $\mu$m together with a strong
IR continuum but only a weak carbonaceous emission at 11.3 $\mu$m is seen; there
is a complete lack of the other AIBs at $\sim$6.2, 7.7, and 8.6 $\mu$m. In this
context, the likely incipient bipolar morphology observed in I22023 (Volk et al.
2004) would support the presence of a binary companion. 

On the other hand, the most recent idea to explain the mixed-chemistry
phenomenon is from Guzm\'an-Ram\'{\i}rez et al. (2011). These authors propose a
chemical model able to form hydrocarbon chains in an UV-irradiated dense torus
in order to explain the high detection rate of mixed-chemistry in PNe of the
Galactic Bulge. However, the UV radiation field in I22023 (T$_{eff}$=24,000 K)
is lower than that in double-dust chemistry PNe (with T$_{eff}$$>$34,000 K) and
may be not intense enough to efficiently break the CO molecules. In addition,
the Spitzer infrared spectrum of I22023 is very peculiar because the O-rich
silicate dust is mostly amorphous and there is no clear evidence for the
presence of crystalline silicate features at wavelengths longer than 20 $\mu$m.
This is in strong contrast with the Spitzer spectra of double-dust chemistry PNe
(e.g., Perea-Calder\'on et al. 2009; Guzm\'an-Ram\'{\i}rez et al. 2011) where
only crystalline silicates are detected.

An alternative explanation to explain the presence of carbonaceous molecules in
I22023 may be non-equilibrium chemistry induced by shocks (Cherchneff 2011).
Cherchneff (2011) demonstrates that water can form in C-rich evolved stars,
showing that, independently of the stellar C/O ratio, thermal fragmentation of
CO occurs in the hot post-shock gas. Our optical spectrum of I22023 shows clear
evidences of on-going mass loss - i.e., the presence of a strong and variable
stellar wind and shocks - which would support this carbonaceous molecules
formation scenario. Indeed, other hot (B-type) post-AGB stars such as IRAS
20462$+$3416 and IRAS 19336-0400 are infrared spectroscopic twins of I22023,
showing both an identical (mixed-chemistry) Spitzer spectrum (see Cerrigone et
al. 2009) together with clear indications (e.g., P-cygni profiles) of on-going
(and variable) mass loss (see Sanchez-Contreras et al. 2008; Arkhipova et al.
2011). In this scenario, the lack of strong infrared features from carbonaceous
molecules in other hot and O-rich post-AGB stars such as IRAS 18062$+$2410 (or
even the very young PN Hen 3-1357)\footnote{No P-cygni profiles (i.e.,
strong stellar winds) are present in IRAS 18062$+$2410 (Arkhipova et al. 2007)
and  based on the C IV 1550 \AA~line in the IUE UV spectrum, the fast wind in
Hen 3-1357 was stopped in 1995 (Parthasarathy et al. 1995).} would be related
with the inactivity of strong stellar winds with significant mass loss rates;
i.e., the absence of strong shocks activating non-equilibrium chemistry. 

In summary, we speculate that the simultaneous presence of carbonaceous
molecules and amorphous silicates in I22023 and other hot (B-type) post-AGB
stars may point to a binary central system with a dusty disk/torus as the
stellar origin common to the hot post-AGB stars hosting O-rich central stars.
The episodic character of the stellar wind (shocks) and mass loss in these hot
O-rich post-AGB stars would favor shock-induced non-equilibrium chemistry as the
carbonaceous molecules formation scenario in these O-rich environments. Further
monitoring studies (e.g., monitoring of radial velocity, light variations,
strengths and profiles of emission and absorption lines) of this star and other
hot post-AGB stars are encouraged in order to understand the circumstellar
mixed-chemistry, mass loss rate (and evolution) with the ultimate goal of
unveiling the stellar origin of this intriguing class of O-rich post-AGB
objects.

\section*{Acknowledgments}

GS would like to acknowledge financial support from the Department of Science 
and Technology (DST), Govt. of India through a grant numbered
SR/FTP/PS$-$67/2005 . D.A.G.H and A.M. also acknowledge support for this work
provided by the Spanish Ministry of Science and Innovation (MICINN) under a JdC
grant and under grant AYA-2007-64748. MP is very thankful to Prof. Shoken Miyama
for his kind support, encouragement and hospitality.

\appendix
\begin{figure*}
\begin{center}
\section*{High resolution optical spectrum of IRAS 22023+5249}
\end{center}
\renewcommand{\thefigure}{4}
\epsfig{figure=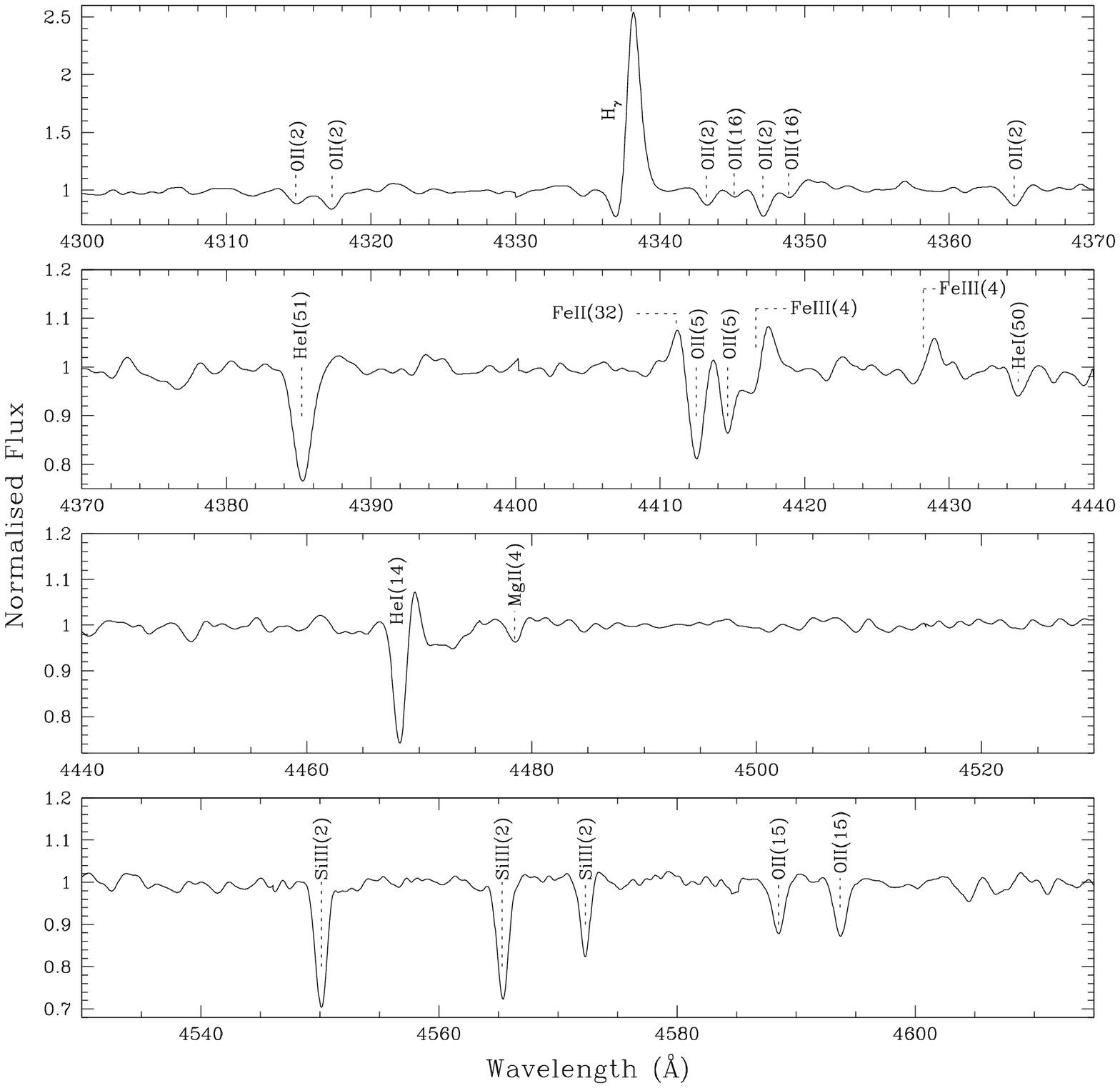, width=18cm, height=23cm}
\caption{Optical spectrum of IRAS22023+5249 (LS~III~+52$^{\circ}$24)}
\end{figure*}

\setcounter{figure}{0}
\begin{figure*}
\renewcommand{\thefigure}{4}
\epsfig{figure=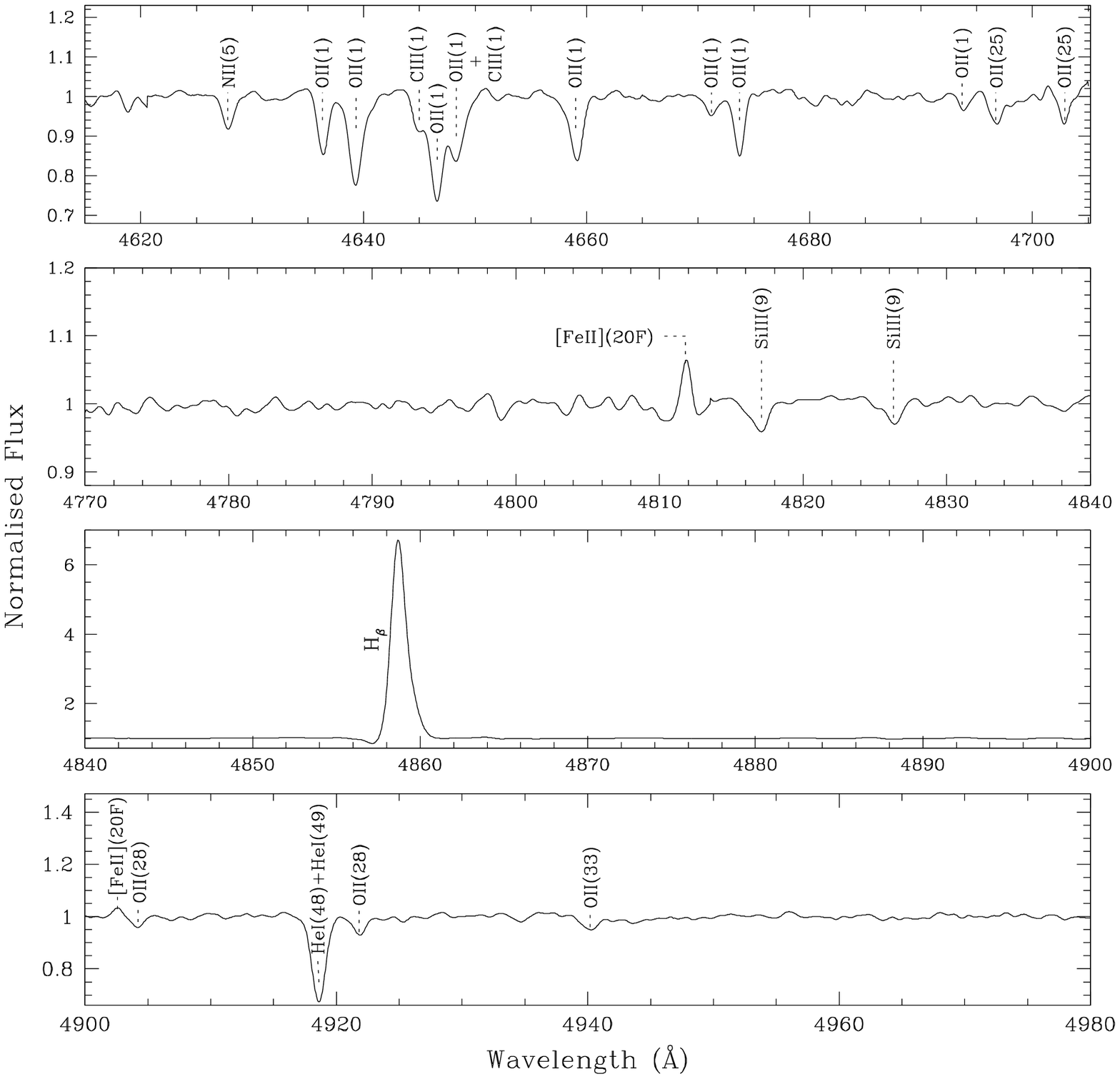, width=18cm, height=23cm}
\caption{Optical spectrum of IRAS22023+5249 (LS~III~+52$^{\circ}$24) contd...}
\end{figure*}

\setcounter{figure}{0}
\begin{figure*}
\renewcommand{\thefigure}{4}
\epsfig{figure=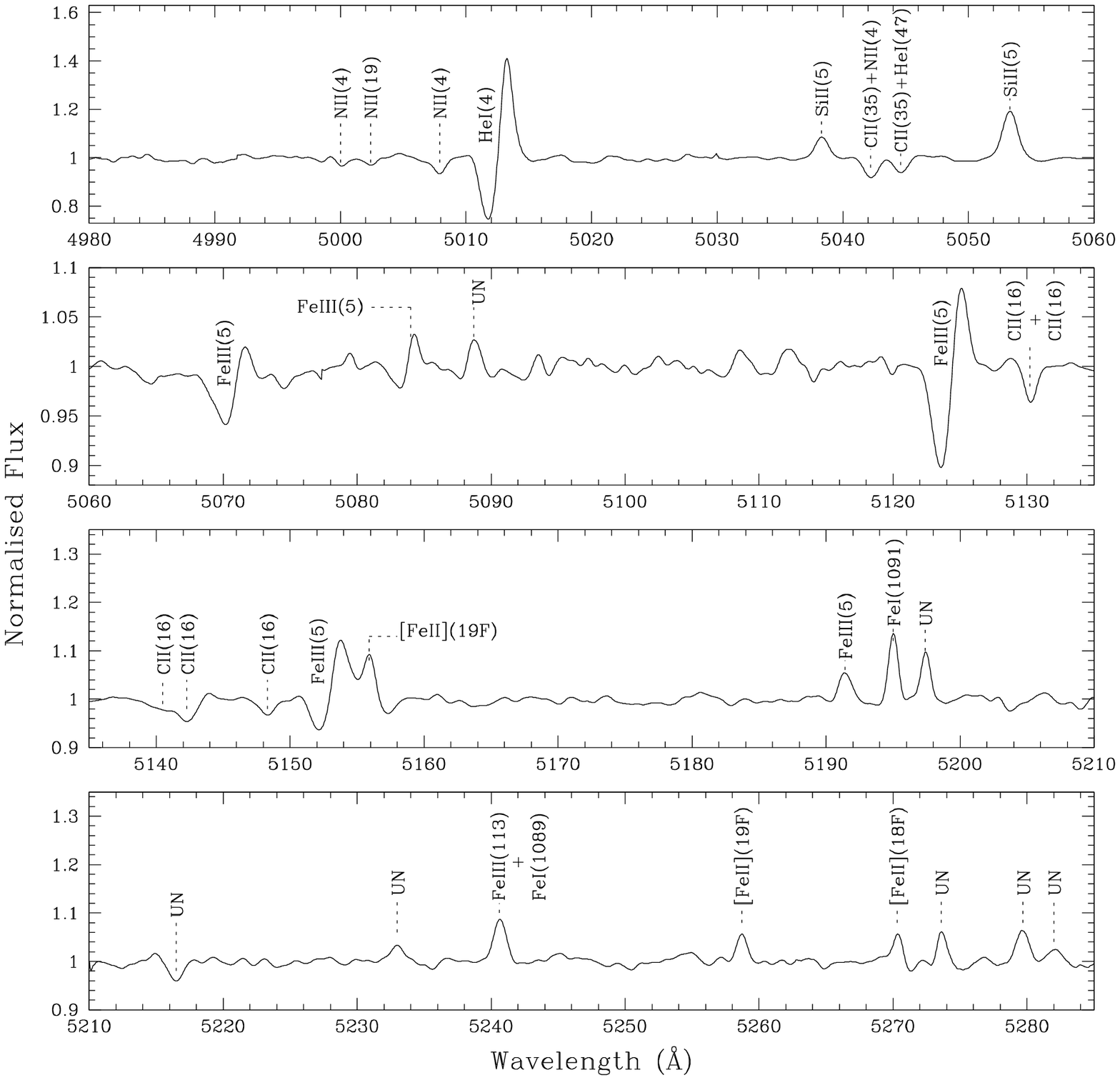, width=18cm, height=23cm}
\caption{Optical spectrum of IRAS22023+5249 (LS~III~+52$^{\circ}$24) contd...}
\end{figure*}

\setcounter{figure}{0}
\begin{figure*}
\renewcommand{\thefigure}{4}
\epsfig{figure=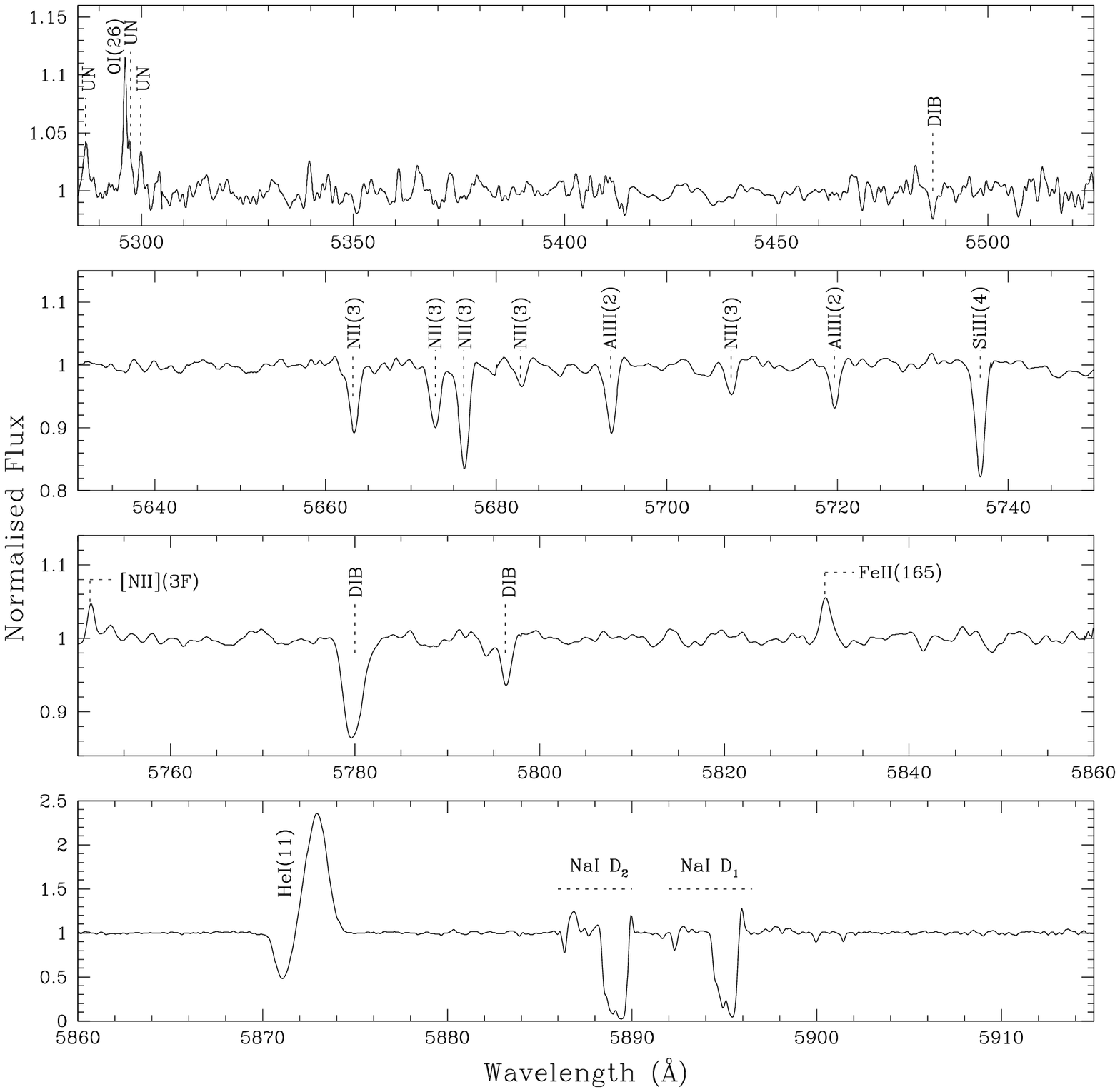, width=18cm, height=23cm}
\caption{Optical spectrum of IRAS22023+5249 (LS~III~+52$^{\circ}$24) contd...}
\end{figure*}

\setcounter{figure}{0}
\begin{figure*}
\renewcommand{\thefigure}{4}
\epsfig{figure=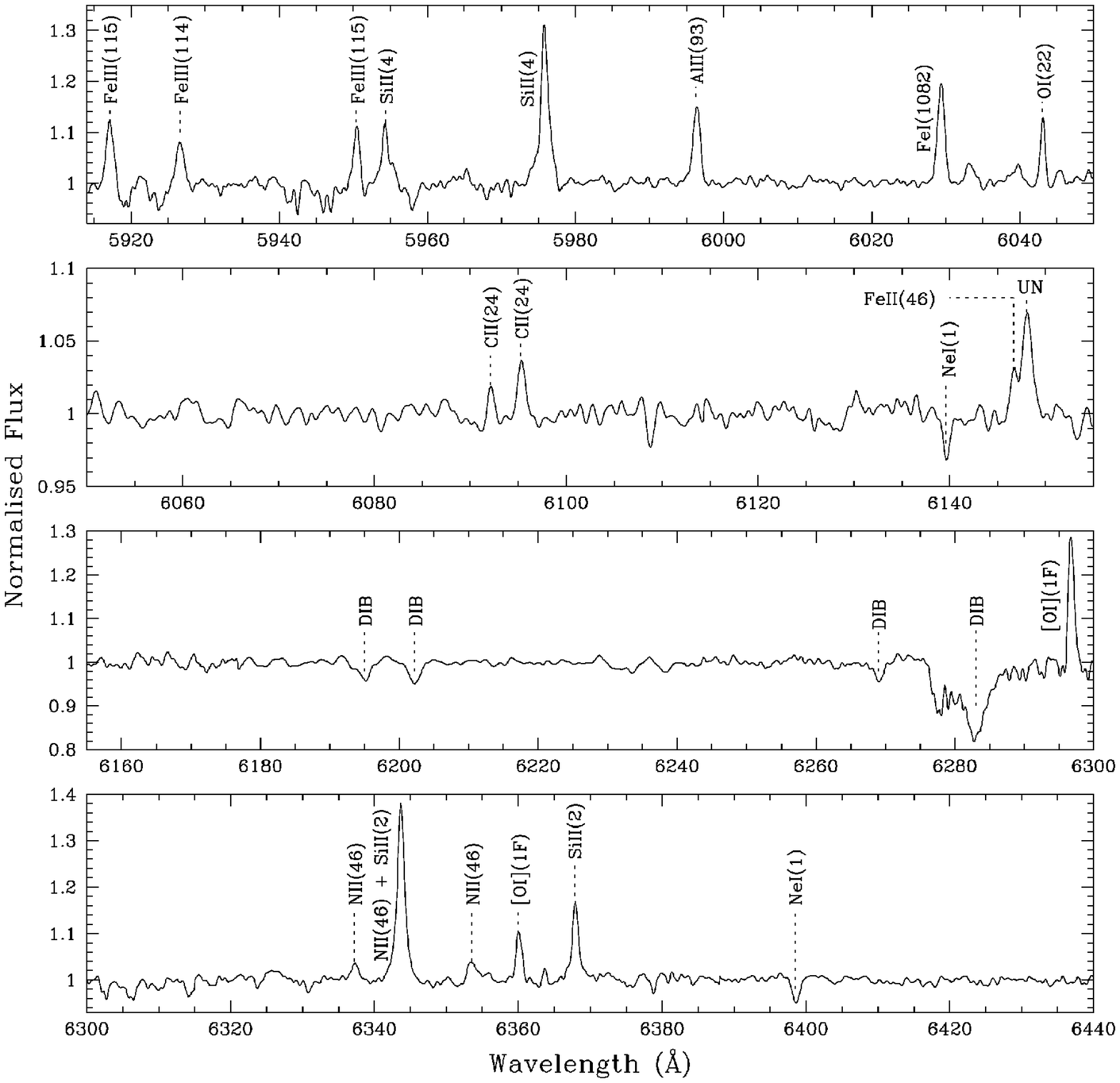, width=18cm, height=23cm}
\caption{Optical spectrum of IRAS22023+5249 (LS~III~+52$^{\circ}$24) contd...}
\end{figure*}

\setcounter{figure}{0}
\begin{figure*}
\renewcommand{\thefigure}{4}
\epsfig{figure=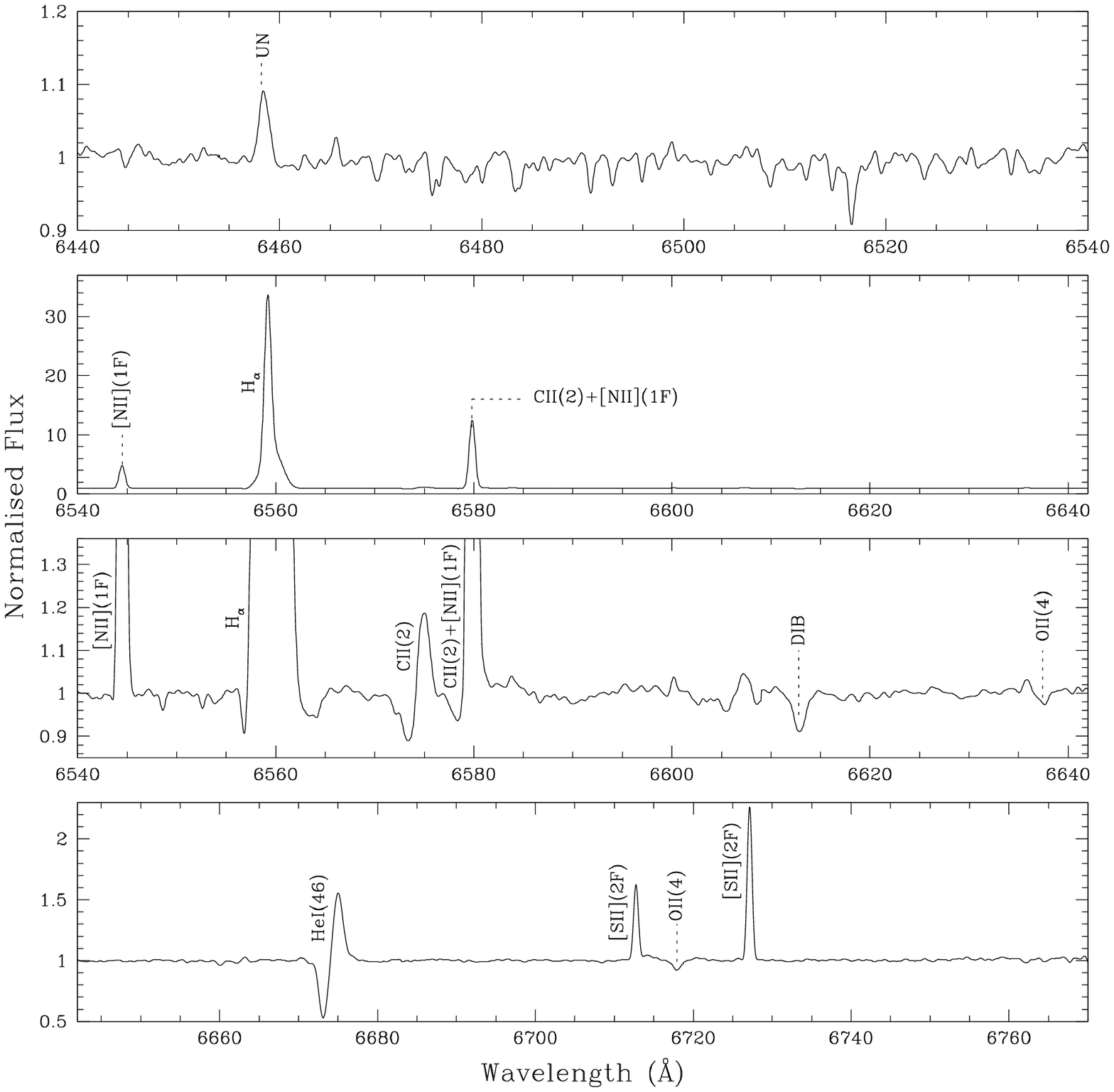, width=18cm, height=23cm}
\caption{Optical spectrum of IRAS22023+5249 (LS~III~+52$^{\circ}$24) contd...}
\end{figure*}

\setcounter{figure}{0}
\begin{figure*}
\renewcommand{\thefigure}{4}
\epsfig{figure=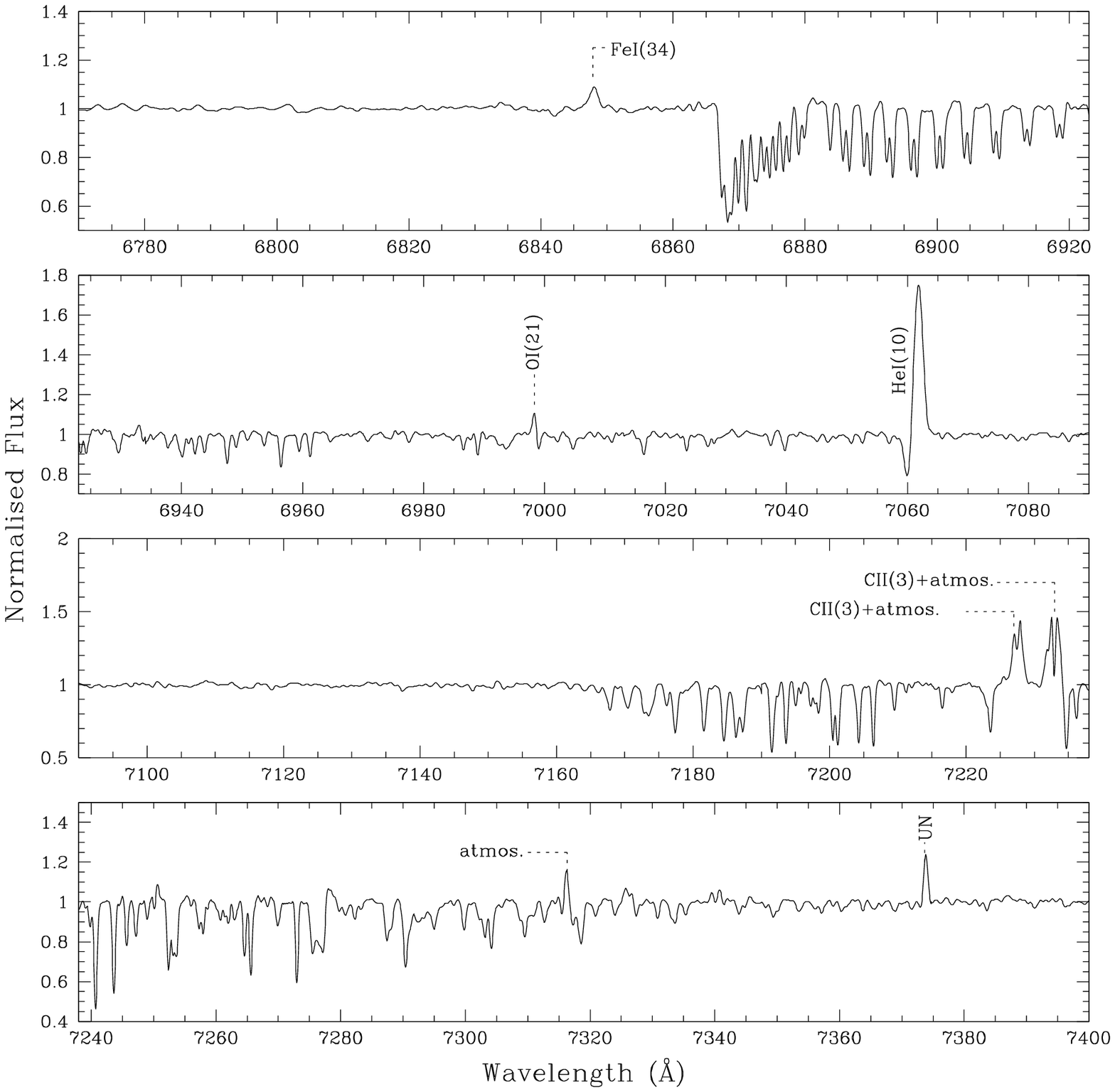, width=18cm, height=23cm}
\caption{Optical spectrum of IRAS22023+5249 (LS~III~+52$^{\circ}$24) contd...}
\end{figure*}

\setcounter{figure}{0}
\begin{figure*}
\renewcommand{\thefigure}{4}
\epsfig{figure=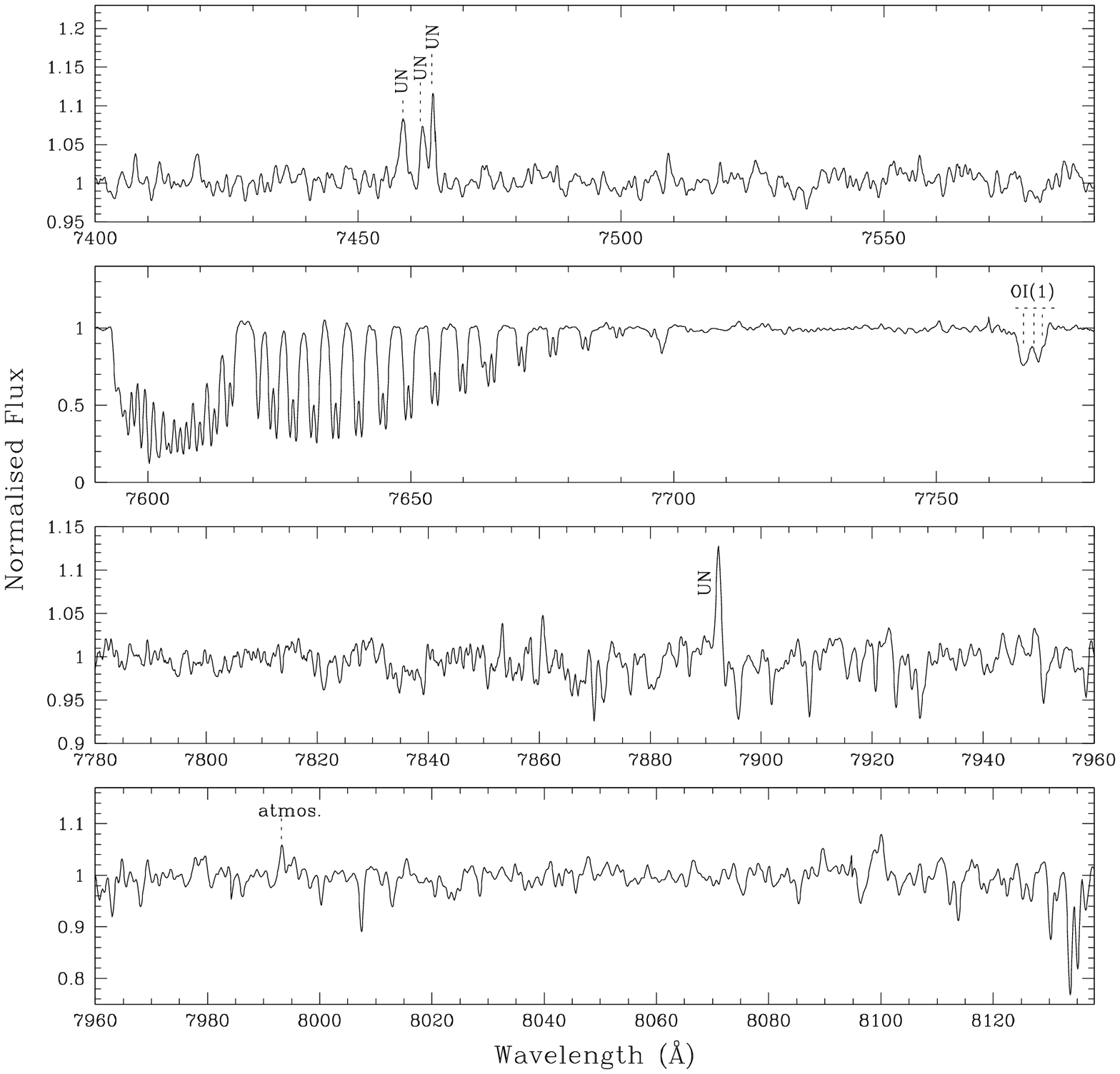, width=18cm, height=23cm}
\caption{Optical spectrum of IRAS22023+5249 (LS~III~+52$^{\circ}$24) contd.
Note that the unmarked absorption line around 7698 \AA~is insterstellar K I.}
\end{figure*}

\setcounter{figure}{0}
\begin{figure*}
\renewcommand{\thefigure}{4}
\epsfig{figure=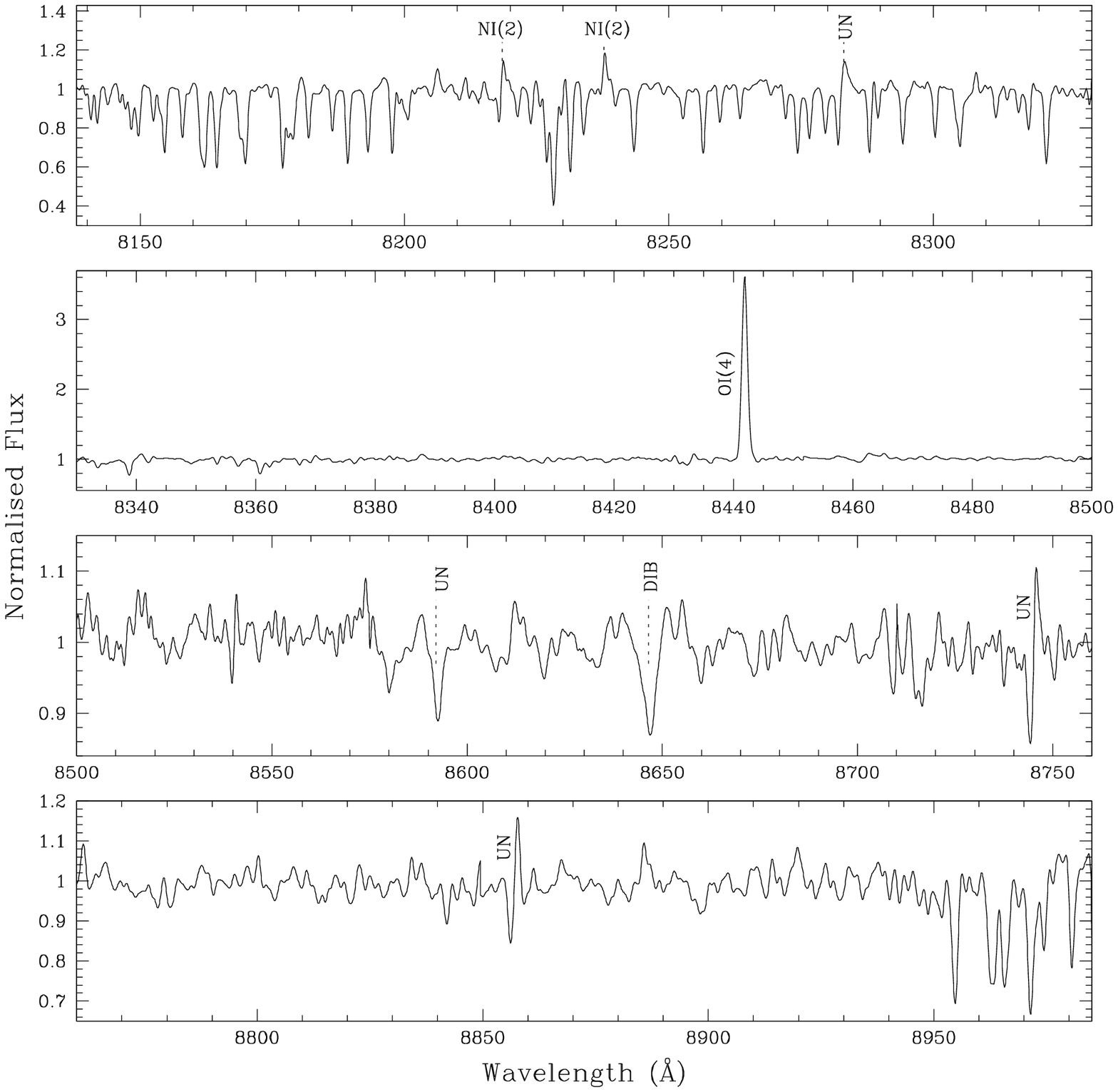, width=18cm, height=23cm}
\caption{Optical spectrum of IRAS22023+5249 (LS~III~+52$^{\circ}$24) contd...}
\end{figure*}

\bsp

\label{lastpage}

\end{document}